\def\d{\mathrm{d}} 
\def\sumpr{\mathop{{\sum}'}} 
\def\tran{{\rm T}}
\def\r#1{\mathbf{r}_{#1}}
\def\rp#1{\mathbf{r}_{#1}^{\prime}} 
\documentclass[aps,twocolumn,showpacs,amsmath,amssymb,pra]{revtex4}
\usepackage{dcolumn} 
\usepackage{bm}

\begin{document} 
\bibliographystyle{apsrev}

\title{Self-consistent calculation of the coupling constant in the
Gross-Pitaevskii equation}
 
\author{A. Yu. Cherny$^{1,2}$ and J. Brand$^1$} 

\affiliation{\mbox{$^1$Max-Planck-Institut f\"ur Physik komplexer
Systeme, N\"othnitzer Stra{\ss}e 38, D-01187 Dresden, Germany}}
\affiliation{$^2$Bogoliubov Laboratory of Theoretical Physics, Joint
Institute for Nuclear Research, 141980, Dubna, Moscow region, Russia}

\date{\today}

\begin{abstract} 
A method is proposed for a self-consistent evaluation of the coupling
constant in the Gross-Pitaevskii equation without involving a
pseudopotential replacement.  A renormalization of the coupling
constant occurs due to medium effects and the trapping potential,
e.g.\  in quasi-1D or quasi-2D systems.  It is shown
that a simplified version of the Hartree-Fock-Bogoliubov approximation
leads to a variational problem for both the condensate and a two-body
wave function describing the behaviour of a pair of bosons in the
Bose-Einstein condensate.
The resulting coupled equations are free of unphysical divergences.
Particular cases of this
scheme that admit analytical estimations are considered and compared
to the literature. In addition to the well-known cases of
low-dimensional trapping, cross-over regimes can be studied.
The values of the kinetic, interaction, external, and release energies
in low dimensions are also evaluated and contributions due to short-range
correlations are found to be substantial.
\end{abstract} 
\pacs{03.75.Hh, 05.30.Jp, 67.40.Db} 
\maketitle 

\section{Introduction} 
\label{sec:intro}

The Gross-Pitaevskii (GP) equation~\cite{gp} is a powerful tool for
describing most of the physical properties of Bose-Einstein
condensates of  trapped alkali atoms
(see reviews~\cite{walls,dalfovo,leggett,pethick}). In the GP approach, the
ground state energy of a trapped dilute Bose gas of atoms of mass $m$
is the functional 
\begin{equation} 
E = \int \d{\bf r}\left[\frac{\hslash^2}{2m}\,|\nabla \phi|^2+
V_{\rm ext}({\bf r}) |\phi|^2 +\frac{g}{2} |\phi|^4 \right], 
\label{gpfun} 
\end{equation} 
of the order parameter $\phi=\phi({\bf r})=\langle \hat\Psi({\bf r})
\rangle$, where $\hat\Psi({\bf r})$ is the bosonic field operator and
$V_{\rm ext}({\bf r})$ is an external trapping potential.
The coupling constant $g$ in the GP functional
(\ref{gpfun}) is intimately related to the density expansion of the
energy of the homogeneous Bose gas. Indeed, in the homogeneous case
Eq.~(\ref{gpfun}) yields $E/N=gn/2$ where $n=N/V\simeq|\phi|^2$
is the 3D density. This expression should be equal to the known first
term in the density expansion $E/N =2\pi\hslash^2a n/m$ in three
dimensions~\cite{bog47,lee}. With $a$ being the 3D scattering length,
the coupling constant $g =4\pi\hslash^2 a/m$ coincides with the zero
momentum limit of the scattering amplitude, the two-body $T$-matrix, for
two particles scattering in a vacuum.  This standard approach based on
the low-density expansion of the homogeneous gas neglects the
influence of inhomogeneous trapping potentials which may require a
renormalization of the coupling constant.

The situation is more complicated for two dimensional Bose gases,
which can be regarded as the limiting case of a 3D gas with a highly
inhomogeneous trapping potential. 
Kolomeisky {\it et al.}~\cite{kolom} proposed that the form (\ref{gpfun})
of the energy functional is still valid. However, in this case the
coupling constant becomes dependent on the local density. Indeed,
Schick's result for the energy $E/N=2\pi\hslash^{2}n_{\rm 2D}/[-m\ln(n_{\rm
2D}a_{\rm 2D}^{2})]$ of a dilute 2D Bose gas~\cite{schick} leads
to the coupling constant $g=4\pi\hslash^{2}/(-m\ln |\phi|^{2}a_{\rm
2D}^{2})$~\cite{kolom} and further corrections were derived in
Refs.~\cite{popov,kolom1,ovchin,cherny4}.  Here, $n_{\rm 2D}$ and $a_{\rm 2D}$ denote
the two dimensional density and scattering length, respectively.  
This generalization can be understood as a local density approximation,
which yields consistent energy values in the homogeneous and
inhomogeneous cases but does not reveal the nature of the additional
non-linearity in the GP equation. Moreover, it is not clear what the
nonlinearity should be in the crossover regimes from 2D to 3D and from 1D to 3D.
A mathematically rigorous justification of the GP
functional~\cite{lieb} is of importance but hardly can help us in this
situation.

The purpose of the present paper is to show how a density-dependent
renormalized coupling constant emerges naturally in a simplified
Hartree-Fock-Bogoliubov (HFB) approximation starting from the bare
interaction potential $V(r)$. To this end, we derive
generalized GP equations where the order parameter is coupled to the
pair wave function of two bosons in the
condensate. The latter has already been discussed in detail in
Refs.~\cite{cherny,cherny2,cherny4,cherny1}.  
The generalized GP equations permit us to consider
interaction potentials with a hard core directly
without the $\delta$-function replacement by accurately accounting
for the short-range spatial correlations of the particles. These
correlations become essential in low dimensions since the Born
approximation for two scattering particles fails at small
momenta (see, e.g., Ref.~\cite{landau}, Sec.~45).
We note that the correct treatment of the short-range correlations
is also possible within the Jastrow\cite{cowell}
and the Faddeev-Yakubovsky~\cite{sorensen} approaches.
 
It is well known that the original HFB scheme leads to an artificial gap in the
spectrum~\cite{hohmartin,Hugenholtz1959a}. Moreover, this scheme in conjunction with the
$\delta$-function replacement has an ultraviolet 
divergence~\cite{lee,leggett1}. These problems then have to
be cured by further approximations as classified by
Griffin~\cite{griffin}. Alternatively, complicated renormalization
procedures \cite{Bijlsma1997a,Hutchinson1998a} or
pseudopotentials~\cite{huang57,olshanii:010402} have been
suggested. In this paper we will discuss a novel approximation derived
from the full HFB scheme where the use of the bare two-body potential
provides an implicit renormalization, and the ultraviolet divergences are avoided.
We will discuss the excitation spectrum and show that it is
gapless in a reasonable approximation.

Low dimensional Bose systems are not only of general
theoretical interest but also find the attention of current
experimental exploration~\cite{ketterle,tolra,paredes04,moritz:250402}.
In these experiments, the low dimensional condensate is
realized by highly anisotropic 3D trapping potentials when the
single-particle energy-level spacing in the tightly confined
dimensions exceeds the interaction energy between atoms
$\hslash\omega_{\rho,z} \gtrsim \mu$. Here the frequencies are
associated with the axially symmetric harmonic potential, and $\mu$ is
the part of the chemical potential coming from the particle
interaction,
which is of order of the mean interaction
energy per particle. This criterion takes the form
$l_{\rho,z}\lesssim \xi$ in terms of the coherence length
$\xi=\hslash/\sqrt{\mu m}$~\cite{note1b}
and the radial (axial) harmonic oscillator
length $l_{\rho,z}=\sqrt{\hslash/(m\omega_{\rho,z})}$. 

Theoretically, the 2D regime $l_z\ll\xi$, $l_\rho\gg\xi$ was
investigated in detail by Petrov, Holzmann, and
Shlyapnikov~\cite{shlyap2D}. The coupling constant was assumed to be
the $T$-matrix of two particles scattering in the harmonic trap with
$l_\rho=\infty$ at the energy of relative motion $E=2\mu$.
An additional nonlinearity is introduced, since the local value of 
$\mu$ depends on  the density and the coupling constant in a self-consistent way.
In this
regime, the motion of particles is confined in $z$-direction to
zero-point oscillation. This implies that the order parameter can be
represented in the form $\phi(x,y,z)=\phi_0(z)\phi(x,y)$, where
$\phi_0(z)$ is the ground state of the 1D harmonic oscillator and
$\phi(x,y)$ is governed by the two dimensional GP equation resulting
from the functional~(\ref{gpfun}) in two dimensions. So, in this
regime, the behaviour of the condensate in $x$-$y$ plane is the same
as in the ``pure'' 2D case with the 2D scattering length written in
terms of the length $l_z$ of the tight confinement~\cite{shlyap2D}.

An improved many body $T$-matrix theory was developed by Stoof and
coworkers~\cite{stoof} in order to describe not only the homogeneous
low-dimensional Bose gases but also the crossover from 3D to lower
dimensions. The coupling constant in the inhomogeneous case is
represented by the local value of the $T$-matrix, which depends on the
local value of the chemical potential. The local $T$-matrix
approximation was also used in Ref.~\cite{burnett} and the microscopic
approach of Ref.~\cite{burnett1}. Thus, one can say, slightly
simplifying the situation, that the common method of evaluating of the
coupling constant in the above works is to determine first the
$T$-matrix from the corresponding two-body Schr\"odinger equation
supposing that the motion of the particles is infinite in some
directions, and after that to replace the coupling constant by the
local value of the $T$-matrix. In this paper we offer a method beyond
the local $T$-matrix approximation.  The coupling constant is
determined self-consistently for a given 3D geometry from a unified
variational scheme. As a result, we obtain a non-local term in the
energy functional, which can be of practical importance if the
external potential
varies on the scale of the interaction potential. This may be
realized, e.g., for condensates of loosely-bound molecules in
tight or strongly oscillating potentials like optical lattices. We
expect experiments to enter this regime in the near future as both
atomic condensates in optical lattices \cite{tolra,Greiner2002a} and
molecular condensates \cite{jochim03,zwierlein:250401,greiner04} are
currently under intense experimental investigation.

As a starting point of our approach we assume that we have a
Bose-Einstein condensate or quasi-condensate with a well-defined order
parameter.
Long-range fluctuations of the phase, which become important for many
physical properties in low-dimensional traps \cite{dettmer01}, are
beyond the scope of our scheme. They can be studied by means of the
approaches of Refs.~\cite{shlyap2D,stoof,shlyap1D,ho}.
Also the strongly-interacting fermionized regime of the
Tonks-Girardeau gas \cite{paredes04}, which was studied in
Ref.~\cite{brand04a}, cannot 
be described with the methods of this paper. 
However, we
note that our scheme, within its accuracy, is simple and physically
transparent and able to reproduce not only the value of the coupling
constant in 1D~\cite{olshanii} and 2D~\cite{shlyap2D} regimes but also
to describe the 3D--2D and 3D--1D crossovers. Furthermore, it allows
us to calculate directly the correct values of the kinetic and
interaction energies of bosons in the trap, which are {\it not} given
by the first and the third terms, respectively, in the GP
functional~(\ref{gpfun})~\cite{cherny3}.
 
The paper is organized as follows. In Sec.~\ref{sec:gpgen} we derive the
generalized GP functional and corresponding equations from a simplified
HFB approximation.  In Sec.~\ref{sec:cross}, a few specific cases are
considered that admit analytical estimations, including the
homogeneous and inhomogeneous Bose gases in low dimensions. In
Sec.~\ref{sec:kinpot} we calculate the values of various contributions
in the energy. In particular, the release energy of the low
dimensional gases is estimated. In Sec.~\ref{sec:vir} we
derive a useful virial theorem and a relation between the chemical
potential and different parts of the energy functional.  The
eigenfunctions of the two-body density matrix and a relation between
the normal and anomalous averages are obtained within the HFB
approximation in Appendices \ref{sec:pwf} and \ref{sec:hfrel},
respectively. 

\section{Generalized Gross-Pitaevskii equations} 
\label{sec:gpgen} 

\subsection{Failure of standard GP approach} 
In the standard approach~\cite{leggett,dalfovo,walls,pethick}, the equilibrium
value of the order parameter $\phi$ is determined by minimization of
the GP functional (\ref{gpfun}) with the constraint of particle-number
conservation $\delta(E-\mu' N) /\delta\phi^{*}({\bf r})=0$. Here
$N\simeq N_0$ is the number of particles and the chemical potential
$\mu'$ appears as a Lagrange multiplier.  Introducing for later
convenience $\mu=\mu'-E_0$ as the chemical potential due to
interaction where $E_0$ is the ground state energy of a
non-interacting particle, we arrive at
\begin{equation} 
(E_0+\mu)\phi=-\frac{\hslash^2}{2m}\nabla^{2}\phi+V_{\rm ext}({\bf r})\phi + g|\phi|^2\phi. 
\label{gpeq} 
\end{equation} 

The simplicity of
this derivation is based on the simple form of the GP energy functional
(\ref{gpfun}), where the effects of the binary inter-particle
interactions has been reduced to a single parameter given by the
coupling constant.
In order to determine the constant
self-consistently, it should be examined carefully how the
interaction term $(g/2)|\phi|^4$ appears in the GP functional
(\ref{gpfun}).
 
In a general many-body system with binary interactions,
the expectation value of the interaction energy
is a functional of the {\it two-body} density matrix
$\langle{\hat\Psi}^{\dagger}(x_1){\hat\Psi}^{\dagger}(x_{2})
{\hat\Psi}(x'_{2}){\hat\Psi}(x'_1)\rangle$. For a  pairwise
interaction potential $V(x_1,x_2) =V({\bf r}_1- {\bf r}_2,
\sigma_1,\sigma_2)$ we thus obtain \cite{bogbog,landau}
\begin{eqnarray} 
E_{\rm int}&=&\Bigl\langle\frac{1}{2}\sum_{i\not=j} V(x_i,x_j)\Bigr\rangle 
=\frac{1}{2}\int\d x_1\d x_2\,V(x_1,x_2)\nonumber \\ 
&&\times\langle{\hat\Psi}^{\dagger}(x_1){\hat\Psi}^{\dagger}(x_{2}) 
{\hat\Psi}(x_{2}){\hat\Psi}(x_1)\rangle, 
\label{eint} 
\end{eqnarray} 
where $x=({\bf r},\sigma)$ stands for the coordinate and spin or
sort indices of a particle, respectively and $\int\d x\cdots =
\sum_{\sigma}\int\d{\bf r}\cdots$.  The kinetic energy and
the energy of interaction with an external field are determined by the
one-body matrix $\langle{\hat\Psi}^\dag(x){\hat\Psi}(x')\rangle$
\begin{eqnarray} 
E_{\rm kin}\!&=&\!\Big\langle\sum_{i}\frac{p_{i}^{2}}{2m}\Big\rangle \nonumber\\ 
\!&=\!&-\frac{\hslash^2}{2m}\int\d x\,\left.\nabla^2_x\langle{\hat\Psi}^\dag(x'){\hat\Psi}(x)\rangle\right|_{x'=x},
\label{ekin}\\
E_{\rm ext}\!&=&\!\Big\langle\sum_{i} V_{\rm ext}(x_{i})\Big\rangle 
\!=\!\int\! \d x\,V_{\rm ext}(x)\langle{\hat\Psi}^\dag(x){\hat\Psi}(x)\rangle. 
\label{eext}  
\end{eqnarray} 
The behaviour of the one- and two-body matrices is easy to understand
in the dilute limit, when the condensate depletion $(N-N_0)/N$ is
small. We note that the number of bosons in the condensate $N_0$ is
defined as the macroscopic eigenvalue of the one-body density matrix
$\langle \hat{\Psi}^{ \dag} (x) \hat{\Psi}(x')\rangle$, that is $\int
\d x'\,\langle \hat{\Psi}^{ \dag} (x') \hat{\Psi}(x)\rangle
\phi_{0}(x') = N_{0} \phi_{0}(x)$.
The field operator can be expanded in the
complete orthonormal set of eigenfunctions of the one-body matrix
$\hat{\Psi}(x)=\hat{a}_0\phi_0(x)+\sumpr_\nu\hat{a}_\nu\phi_{\nu}(x)$,
where the sum $\sumpr_\nu$ means $\sum_{\nu\not=0}$ and $\int \d
x\,|\phi_{\nu}(x)|^{2}=1$. Appearance of the 
Bose-Einstein condensate implies the macroscopic occupation of $N_0$,
i.e.\ the ratio $N_0/N$ remains finite in the thermodynamic limit.
Following Bogoliubov \cite{bog47,bogquasi} we now replace the
condensate operators by $c$-numbers
$\hat{a}^\dag_{0}=\hat{a}_{0}\simeq \sqrt{N_0}$ 
and represent the Bose field operator in the form $\hat{\Psi}(x)=
\phi(x) +\hat{\vartheta}(x)$~\cite{castin98n}. Here $\phi(x)=
\sqrt{N_0}\phi_0(x)$ is the $c$-number part, and $\hat{\vartheta}(x)
=\sumpr_\nu\, \hat{a}_\nu \phi_\nu(x)$ the operator part, for which we
have $\langle\hat{\Psi}(x)\rangle =\phi(x)$ and
$\langle\hat{\vartheta}(x)\rangle=0$. Thus, the order parameter is
nothing else but the non-normalized eigenfunction of the one-body
density matrix.
 
In the original approach of Gross and Pitaevskii, the simplest
mean-field approximation is used when the operator part is completely
neglected: $\hat{\Psi}(x) \simeq\phi(x)$ and $\hat{\Psi}^\dag(x)
\simeq\phi^*(x)$. Assuming additionally  that the order
parameter does not change significantly at the distances of order of
the radius $R_{\rm e}$ of the interaction potential, we obtain the 
GP energy functional (\ref{gpfun}) for spinless bosons from
Eqs.~(\ref{eint}--\ref{eext}) with the coupling constant
\begin{equation} \label{eqn:couplingc}
g\simeq g_{\rm B}\equiv\int\d{\bf r}\,V(r).
\end{equation}
This coupling constant can be identified with the two-body scattering
amplitude at zero momentum in the Born approximation.  The validity of
the GP approach with the coupling constant $g_{\rm B}$ of
Eq.~(\ref{eqn:couplingc}) is certainly linked to the validity
condition of the Born approximation at zero momentum that the
potential $V(r)$ be small and integrable.

Of the two assumptions mentioned above, namely the slow spatial
variation of the order parameter and the validity of the Born
approximation, the former is usually fulfilled as the healing length
$\xi=\hslash/\sqrt{\mu m}$ as a lower bound of the length scale of the
order parameter \cite{dalfovo,leggett} is usually much larger than the
effective range of the interaction. The latter assumption, however, is
clearly not fulfilled for experiments with cold atomic gases as their
interactions are of the hard-core type.

The validity of the GP approach can be extended to such systems by an
argument attributed to Landau~\cite{bog47,cherny1}. He noted that at extremely low
energies, as predominant in the dilute-gas BEC, the scattering
properties are completely determined by only one single parameter,
which is the 3D $s$-wave scattering length $a$.  This allows us
to replace the Born approximation for the scattering amplitude $g_{\rm
B}$ by its exact value $g=4\pi\hslash^2a/m$, which can be found from
the two-body Schr\"odinger equation even for hard-core potentials.
This indirect argument, however, cannot be used in one or two
dimensions as there is no such simple relation between the integral in
Eq.~(\ref{eqn:couplingc}) and the scattering amplitude as we have in
three dimensions. Furthermore the Born series for the scattering
amplitude diverges for small momenta in two dimensions and below (see, e.g.,
Ref.~\cite{landau}, Sec.~45).

\subsection{Pair wavefunction in a medium}

The above described deficiencies of the naive GP approach may be
remedied by accounting for the two-particle scattering processes
explicitely.  Within the HFB scheme this is possible through certain
correlations introduced by the fluctuation operators
$\hat{\vartheta}$.  In order to see the relation between the
two-particle scattering process and the correlation functions
mentioned above it is useful to introduce the concept of a two-body or
{\em pair wave function in the medium} of other
particles~\cite{bogquasi,cherny}. The pair wave functions in the
medium of the many-body system are defined as eigenfunctions of the
two-body density matrix, as discussed in detail in Appendix
\ref{sec:pwf}. They should be distinguished from the two-body wave
functions in the vacuum, which relate to a system of two
particles. For the latter we will use the superscript $^{(0)}$. 

Let us suppose that we know the exact eigenfunctions of the two-body
density matrix. Then we can expect for the dilute gas, where
low-momentum two-body processes dominate the behaviour of the system,
that the pair wave functions in the medium should be very close~\cite{cherny} to the
two-body wave functions in the vacuum, which are the solutions of the
two-body Schr\"odinger equation.
This physical assumption was used to obtain the density expansions for
the 3D~\cite{cherny1,cherny2} and 2D~\cite{cherny4} homogeneous Bose
gases in a very simple manner.
However, various approximations in the many-body theory can break this
relation.  

\subsection{A simplified HFB scheme}

Within the HFB approximation for the homogeneous Bose gas,
all eigenfunctions of the two-body density matrix except for one are
plane waves and are thus treated in the Born approximation as
shown in Appendix \ref{sec:pwf}. This is an obvious drawback of the
HFB scheme. It turns out that the two-body wave function that is not a
plane wave is proportional to the anomalous average
\begin{equation}
 \varphi(x_1,x_2)\equiv\langle{\hat\Psi}(x_1) {\hat\Psi}(x_{2})\rangle
\end{equation}
and
corresponds to the macroscopic eigenvalue $N_0(N_0-1)$ in the limit of
large $N$. It is the pair wave function that  describes the
two-particle scattering 
process in the medium of the
Bose-Einstein condensate~\cite{cherny,note1}.
Thus we can go beyond the Born
approximation in the framework of the HFB method by keeping only the
contribution of the anomalous average $\langle{\hat\Psi}(x_1)
{\hat\Psi}(x_{2})\rangle$ and neglecting the contribution of the
other wave functions in the two-body density matrix. Due to small
condensate depletion $(N-N_0)/N\ll 1$, one can expect that the
contribution of only this eigenfunction will be sufficient for
obtaining the coupling constant in the GP equation. In
this simplified version of the HFB approximation we set
\begin{equation} 
\langle{\hat\Psi}^{\dagger}(x_1) {\hat\Psi}^{\dagger}(x_{2}) {\hat\Psi}(x'_{2}) {\hat\Psi}(x'_1)\rangle\simeq 
\varphi^*(x_1,x_2)\varphi(x'_1,x'_2). 
\label{2matr} 
\end{equation} 
Extracting the $c$-number part of the field operator, the anomalous
averages can be rewritten as
\begin{equation} 
\varphi(x_1,x_2)=
\phi(x_1)\phi(x_2)+\psi(x_1,x_2), 
\label{phidef} 
\end{equation} 
where we introduced the notation $\psi(x_1,x_2) \equiv
\langle\hat{\vartheta}(x_1) \hat{\vartheta}(x_2)\rangle$ for the
anomalous two-boson correlation function associated with the
scattering part of the two-body wave function. The functions
$\varphi(x_1,x_2)$ and $\psi(x_1,x_2)$ are symmetric with respect to
permutation of $x_1$ and $x_2$ due to the commutation relations for
the Bose field operators. 

For the one-body density matrix we find
$\langle{\hat\Psi}^\dag(x){\hat\Psi}(x')\rangle= \phi^*(x)\phi(x')
+\langle\hat{\vartheta}^\dag(x) \hat{\vartheta}(x')\rangle$. We
note that the normal $\langle\hat{\vartheta}^\dag(x)
\hat{\vartheta}(x')\rangle$ and anomalous $\langle\hat{\vartheta}(x)
\hat{\vartheta}(x')\rangle$ averages are not independent quantities
as discussed in  Appendix~\ref{sec:hfrel} and
Refs.~\cite{cherny1,cherny2}. Within the Hartree-Fock-Bogoliubov
scheme, the relations between them
appear as a specific property of the HFB
ground state (the quasiparticle vacuum) and do not contain parameters
of the Hamiltonian in explicit form.
We will use the approximate relation (\ref{thetaapp}), which
leads to
\begin{equation} 
\langle{\hat\Psi}^\dag(x){\hat\Psi}(x')\rangle= \phi^*(x)\phi(x') 
+\int \d x_2\,\psi^*(x,x_2)\psi(x_2,x'). 
\label{onebody} 
\end{equation} 
With the help of Eqs.~(\ref{2matr}), (\ref{phidef}), and (\ref{onebody}) we rewrite Eqs.~(\ref{eint}), (\ref{ekin}), 
and (\ref{eext}) in terms of the anomalous averages 
\begin{eqnarray} 
E_{\rm int}&=&\frac{1}{2}\int \d x_1\d x_2\,V(x_1,x_2)|\varphi(x_1,x_2)|^2,
\label{eint1}\\ 
E_{\rm ext}&=&\frac{1}{2}\int \d x_1\d x_2\,[V_{\rm ext}(x_1)+V_{\rm ext}(x_2)]|\psi(x_1,x_2)|^2\nonumber\\ 
&&+\int \d x_1\,V_{\rm ext}(x_1)|\phi(x_1)|^2,
\label{eext1}\\ 
E_{\rm kin}&=&\frac{1}{2}\int \d x_1\d x_2\,\psi^*(x_1,x_2)(\hat{T}_1+\hat{T}_2)\psi(x_1,x_2)\nonumber\\ 
&&+\int \d x_1\,\phi^*(x_1)\hat{T}_1\phi(x_1),
\label{ekin1}
\end{eqnarray} 
where $\hat{T}_j=-\hslash^2\nabla^2_{j}/(2m)$ and $j=1,2$. The total
number of particles is related directly to the one-body matrix:
$N=\int\d x\,\langle{\hat\Psi}^\dag(x){\hat\Psi}(x)\rangle$. With
Eq.~(\ref{onebody}) we find
\begin{equation}\label{N} 
N=\int \d x_1\,|\phi(x_1)|^2 + \int \d x_1\d x_2\,|\psi(x_1,x_2)|^2. 
\end{equation} 
The current approximations are useful for a variational scheme where
the functions $\phi(x_1)$ and $\psi(x_1,x_2)$ are determined by
minimization of the total energy with the constraint $N={\rm
const}$. Using the Lagrange method, we obtain the conditions $\delta
E/\delta\phi(x_1)= \delta E/\delta\phi^{*}(x_1) =\delta
E/\delta\psi(x_1,x_2) =\delta E /\delta\psi^*(x_1,x_2) =0$ for the
energy functional
\begin{equation}
E[\{\phi,\psi\},\mu']=E_{\rm kin}+E_{\rm ext}+E_{\rm int}-\mu'(N-{\cal N}),
\label{efunc}
\end{equation}
given by Eqs.~(\ref{eint1})-(\ref{N}). Here $\mu'=\mu +E_0$ is the
chemical potential 
and ${\cal N}=\langle \hat{N}
\rangle$  is the average number of particles, i.e.\ the
l.h.s.\ of Eq.~(\ref{N}) at the equilibrium values of the functions
$\phi$ and $\psi$ corresponding to the minimum (ground state) of the
functional~(\ref{efunc}). Note that the variation
$\delta\psi(x_1,x_2)$ is symmetric under the permutation of $x_1$ and
$x_2$, so, the equation $\int \d x_1\d
x_2\,g(x_1,x_2)\delta\psi(x_1,x_2)=0$ leads to
$g(x_1,x_2)+g(x_2,x_1)=0$ for arbitrary functions $g(x_1,x_2)$. 

This
variational procedure yields the following system of equations for the
one- and two-body functions $\phi(x_1)$ and $\varphi(x_1,x_2)$,
respectively,
\begin{eqnarray} 
\mathcal{L}_1\phi(x_1)+ \int \d x_2\, \phi^{*}(x_2)V(x_1,x_2)\varphi(x_1,x_2)&=&0, 
\label{phieq} \\ 
(\mathcal{L}_1+\mathcal{L}_2)\psi(x_1,x_2) + V(x_1,x_2)\varphi(x_1,x_2)&=&0, 
\label{psieq} 
\end{eqnarray} 
where the operators $\mathcal{L}_1$ and $\mathcal{L}_2$ stand for 
\[ 
\mathcal{L}_j=-\hslash^2\nabla_{j}^2/(2m)-\mu -E_0 + V_{\rm ext}(x_j),\quad j=1,2,
\] 
and $\phi$, $\varphi$ and $\psi$ are simply related by
Eq.~(\ref{phidef}). Due to this relation, Eq.~(\ref{phieq}) is
nonlinear with respect to $\phi$ and can be associated with the GP
equation. Equation (\ref{psieq}) is the analogue of the two-particle
Schr\"odinger equation and is linear with respect to $\varphi$, though
not uniform. The obtained system of two equations allows us to
determine the coupling constant self-consistently. A specific feature
of Eq.~(\ref{phieq}) is the {\it non-local} nature of the last term, which
can play a role when the radius of the interacting potential becomes
of the order of the characteristic length of the anisotropic trapping
potential in some direction, say, $R_{\rm e} \sim l_z\ll \xi$, or if the
trapping potential has a strongly oscillating contribution
with the scale of the order of $R_{\rm e}$. At the same time,
Eqs.~(\ref{phieq}) and (\ref{psieq}) indeed reduce to the GP
equation with the 3D coupling constant $g =4\pi\hslash^2 a/m$ in the
limit $R_{\rm e}\ll \xi \ll l_x,\ l_y,\ l_z$ as will be shown in
Sec.~\ref{sec:gpregime}.

When the external potential becomes independent of some coordinate,
say $z$, particles can move freely in $z$-direction and we should
impose the boundary conditions that follow from Bogoliubov's principle
of correlation weakening~\cite{bogquasi}:
\[
\langle{\hat\Psi}(x){\hat\Psi}(x')\rangle \simeq\langle{\hat\Psi}(x)
\rangle\langle{\hat\Psi}(x')\rangle \quad\text{when}\quad
|z-z'|\to\infty .
\] 
Physically, this implies that the function
$\psi(x_1,x_2) =\langle\hat{\vartheta}(x_1)
\hat{\vartheta}(x_2)\rangle$ vanishes at the spatial distances of
order of the coherence (healing) length, $|{\bf r}_1-{\bf r}_2|\gtrsim
\xi$.

A time-dependent generalization of Eqs.~(\ref{phieq}) and
(\ref{psieq}) can in principle be derived from the equations of motion
of the field operators. Here, however, we will not elaborate the full
derivation but instead present a simple argument that leads to a
useful time-dependent scheme. In the case of a time-independent
Hamiltonian, it can be easily seen that the GP order parameter depends
on time as
\begin{eqnarray}
\phi(x,t)\!&=&\!\langle N-1|{\hat\Psi}(x,t)|N\rangle
\nonumber\\ 
\!&=&\!\phi(x)\exp[{-i(E_{N}-E_{N-1})t}/{\hslash}] \nonumber\\
\!&=&\!\phi(x)\exp[{-i\mu't}/{\hslash}] . \nonumber
\end{eqnarray}
Here, $|N\rangle$ and $E_{N}$ are the ground state and energy of $N$
bosons, respectively. By analogy, we
find $\psi(x_1,x_2,t)=\psi(x_1,x_2)\exp[{-i 2\mu' t/\hslash}]$ and
$\varphi(x_1,x_2,t)=\varphi(x_1,x_2)\exp[{-i 2\mu't/\hslash}]$. We now
argue that the chemical potential in  Eqs.~(\ref{phieq}) and
(\ref{psieq}) arises due to time derivatives, which leads to the
obvious generalization
\begin{eqnarray} 
i\hslash\frac{\partial}{\partial t}\phi(x_1,t)&=&\hat{H}_{1}\phi(x_1,t)+ E_{\rm nl}(x_1,t),
\label{phieqt} \\ 
i\hslash\frac{\partial}{\partial t}\varphi(x_1,x_2,t)&=&[\hat{H}_{1}+\hat{H}_{2}+V(x_1,x_2)]\varphi(x_1,x_2,t)
\nonumber\\&&
+ \phi(x_1,t)E_{\rm nl}(x_2,t)
\nonumber\\&&
+ \phi(x_2,t)E_{\rm nl}(x_1,t), 
\label{psieqt} 
\end{eqnarray}
where we denote
\begin{eqnarray}
\hat{H}_{j}&=&-\frac{\hslash^2\nabla_{j}^2}{2m} + V_{\rm ext}(x_j,t),\quad j=1,2,
\nonumber \\
E_{\rm nl}(x,t)&=&\int \d y\, \phi^{*}(y,t)V(x,y)\varphi(x,y,t).
\nonumber
\end{eqnarray}
The functions $\phi=\langle{\hat\Psi}(x,t)\rangle$ and
$\varphi=\langle{\hat\Psi}(x_{1},t){\hat\Psi}(x_{2},t)\rangle$ are
normalized as $\int\d x\,|\phi(x,t)|^{2} =N_{0}$ and $\int\d
x_{1}\d x_{2}\, |\varphi(x_{1},x_{2},t)|^{2} =N_{0}(N_{0}-1)\simeq
N^{2}_{0}$, respectively.  The time-dependent generalized GP equations
(\ref{phieqt}-\ref{psieqt}) become the ordinary one- and two-body
Schr\"odinger equations, respectively, in the limit $\xi \gg
l_{x},l_{y},l_{z}$ when we can neglect all the nonlinear terms, which
are responsible for many-body effects. Therefore they are of slightly
more general validity than the stationary equations
(\ref{phieq}-\ref{psieq}), which imply a large particle number, since
$E_{N}-E_{N-2}\simeq 2(E_{N}-E_{N-1})\simeq 2\mu'$ is valid only for
$N\gg 1$. We notice that the time-dependent equations similar to that of~(\ref{phieqt}) and (\ref{psieqt})
were derived in papwer~\cite{kohler} by the method of noncommutative cumulants.

\subsection{Properties and limits of validity}

The time-dependent Equations (\ref{phieqt}-\ref{psieqt}) give access
to the elementary excitation spectrum. At the moment we cannot prove the
gaplessness of the spectrum in the
most general case, but we can solve for the excitation energies
approximately. With the ansatz $\varphi({\rm r}_{1},{\rm r}_{2},t)=\phi({\rm
r}_{1},t)\phi({\rm r}_{2},t)[1+\psi(r)/n_0]$, we obtain the Bogoliubov
excitation energy $\omega_{k}=\sqrt{T_k^2+2n_{0}U(k)T_k}$ with the
$k$-dependent scattering amplitude $U(k)$ (for the notations see
Sec.~\ref{sec:hom}). This form of the spectrum for a singular
two-particle interaction was proposed without derivation by Bogoliubov
in Ref.~\cite{bog47}. For small $k$ we can replace $U(k=0) =
4\pi\hbar^2 a/m$ and obtain the usual (gapless) Bogoliubov dispersion.
The additional features in the obtained spectrum at medium and high energies
reflect the structure of the interaction potential neglected in the standard
GP approach and present a clear advantage of our extended scheme. As a consequence,
we can expect that Levinson's theorem for quasi-particle scattering~\cite{brand1}
will be modified.

Let us discuss limits of validity of the generalized GP equations
(\ref{phieq}) and (\ref{psieq}). First, we imply that the
Bose-Einstein condensate (or quasi-condensate in low dimensions, see
Sec.~\ref{sec:nonhom}) is developed strongly. This means that
$r_0\ll\xi$, where $r_0$ is an average distance between
bosons~\cite{dalfovo,leggett}. Second, the above derivation can be
applied only to the short-range interaction potentials that decrease
at least as fast as $V(r)\sim 1/r^{\varepsilon+D}$ for $r\to\infty$,
where $D$ is dimension and $\varepsilon>0$ \cite{cherny1,cherny2}. For
a long-range interaction like Coulomb repulsion, the approximation
(\ref{2matr}) works badly. Third, the approximations (\ref{2matr}) and
(\ref{thetaapp}) are insufficient to describe the long-range behaviour
of the normal $\langle{\hat\vartheta}^{\dag}(x_1)
{\hat\vartheta}(x_{2})\rangle$ and anomalous
$\langle{\hat\vartheta}(x_1){\hat\vartheta}(x_{2})\rangle$ correlation
functions, which are governed by Bogoliubov's ``$1/q^{2}$''
theorem~\cite{bogquasi,mullin,fischer}. According to this theorem, the
above correlation functions should decay as $1/|{\bf r}_1-{\bf
r}_2|^{2}$ when $|{\bf r}_1-{\bf r}_2|\gtrsim \xi$ at zero temperature
if the Bose-Einstein condensate exists. Our scheme gives $1/|{\bf
r}_1-{\bf r}_2|$ decay, as we show in Sec.~\ref{sec:hom}. However, we
stress that {\it the long-range behaviour is not needed for obtaining
the coupling constant}, since the integral in Eq.~(\ref{phieq})
contains the anomalous correlation function multiplied by the short-range
potential $V(x_1,x_2)$ with the characteristic radius $R_{\rm
e}\lesssim \xi$. Since the developed scheme describes well only the
short-range behaviour of $\psi(x_1,x_2)$ for $|{\bf r}_1-{\bf
r}_2|\lesssim \xi$, the integration in the last term of Eq.~(\ref{N})
should be restricted to this region 
\begin{equation}
N=\int \d x_1\,|\phi(x_1)|^2 + \int_{|{\bf r}_1-{\bf
r}_2|\leqslant\xi} \d x_1\d x_2\,|\psi(x_1,x_2)|^2,  
\label{13a}
\end{equation}
otherwise we obtain formally divergent term. 
This modification of the original scheme, however, does not change the
working equations (\ref{phieq} - \ref{psieqt}) 
in the region $|{\bf r}_1-{\bf r}_2|\leqslant \xi$, which is of sole
interest for our purposes. We stress that Eq.~(\ref{13a}) is really
needed only when minimizing the energy functional (\ref{efunc})
directly. Furthermore, if we obtain the solutions of
Eqs.~(\ref{phieq}) and (\ref{psieq}) as functions of the chemical
potential in the grand canonical ensemble, then the condition
(\ref{N}) or (\ref{13a}) can be employed without the second term at
all in order to rewrite the answer in terms of the total number of
particles in the canonical ensemble, owing to small condensate
depletion.

Note that the standard HFB approximation can be obtained by using the
variational scheme if, first, one substitutes
Eqs.~(\ref{eint})-(\ref{eext}) into the energy functional
(\ref{efunc}), second, employ the restrictions Eqs.~(\ref{fphicord1})
and (\ref{fphicord}), and third, retain all additional terms missing
in Eq.~(\ref{2matr}), where the three- and four-boson averages of
$\hat{\vartheta}$ and $\hat{\vartheta}^{\dag}$ ought to be evaluated
by means of the Wick's theorem and, consequently, the three-boson
averages vanish.

\section{Examples} 
\label{sec:cross}

In this section we restrict ourselves to spinless bosons with an
isotropic short-range interaction $V=V(r)$, where $r=|{\bf r}_1 -{\bf
r}_2|$. Even after this simplification, the solution of the generalized
GP equations (\ref{phieq}) and (\ref{psieq}) remains a rather complex
problem. Nevertheless, in a number of specific limiting cases we are
able to obtain analytic results.

\subsection{The homogeneous case} 
\label{sec:hom} 
 
Let us investigate Eqs.~(\ref{phieq}) and (\ref{psieq}) in three and two dimensions for the homogeneous Bose gas. In the
homogeneous case $V_{\rm ext}=0$, hence we have $\psi=\psi(r)$, $E_0=0$, and Eq.~(\ref{phieq}) gives the trivial solution
$\phi=\sqrt{n_0}={\rm const}$. In this subsection, we use the common notation $n_{0}$  for the condensate density in both
2D and 3D cases. Thus, Eqs.~(\ref{phieq}) and (\ref{psieq}) read
\begin{eqnarray} 
\mu&=&\int\d{\bf r}\,V(r)[n_0+\psi(r)], 
\nonumber\\ 
2\mu\psi(r)&=&-\frac{\hslash^2}{m}\nabla^2\psi(r) + V(r)[n_0+\psi(r)],
\nonumber
\end{eqnarray} 
and $\psi(r)\to0$ for $r\to\infty$ in accordance with Bogoliubov's
principle of correlation weakening. Taking the Fourier transformation
of the last equation, we obtain
\begin{eqnarray} 
\mu&=&n_0U(0), 
\label{muhom} \\ 
\frac{\psi(k)}{n_0}&=&-{\rm P.P.}\frac{U(k)}{2(T_k-\mu)}, 
\label{psik} 
\end{eqnarray} 
where we denote $U(k)= \int\d{\bf r}\, V(r)e^{-i{\bf k} \cdot{\bf
r}}[1+\psi(r)/n_0]$, $T_k= \hslash^2k^2/(2m)$, and the symbol
P.P. stands for the principle value of the associated integral. The
latter appears as a natural regularization for the singular
denominator in the r.h.s of Eq.~(\ref{psik}) and implies that the
scattering part of the two-body wave function $\psi(k)$ is real and
corresponds to a standing wave. Note that another regularization, such
as the standard replacement $k\to k\pm i\varepsilon$, leads to the
same results in the leading order at small densities. Within the more
accurate method~\cite{cherny1,cherny2}, we obtain the same equation as
(\ref{psik}) but with the Bogoliubov denominator $2\sqrt{T_k^2+2n_0U(k)
T_k}$. The latter provides the correct values of {\it both} the short-
and long-range behaviour of the correlator
$\psi(r)=\langle\hat{\vartheta}({\bf r}) \hat{\vartheta}(0)\rangle$
[which is the Fourier transform of $\psi(k)$], while Eq.~(\ref{psik})
provides only the short-range behaviour. Indeed, in the 3D case we
have $\psi(r)\sim \cos(\sqrt{2}r/\xi)/r$ at $r\gtrsim\xi$ (see below) but not $\psi(r)\sim 1/r^{2}$ as it should be.

Equation (\ref{psik}) can be  rewritten in the Lippmann-Schwinger form with the help of the Fourier transformation. By
using the familiar property of  Fourier transformation $\int \d{\bf k}\, e^{i{\bf k} \cdot{\bf r}} g({\bf k})f({\bf k})
/(2\pi)^D= \int\d{\bf r}'f({\bf  r}')g({\bf r}-{\bf r}')$ (here $D$ is the dimension), we obtain the equation for
$\varphi(r)=n_0+\psi(r)$ 
\begin{equation} 
\label{phir} 
\varphi(r)=n_0+\int\d{\bf r}'\,V(r')\varphi(r')G(|{\bf r}-{\bf r}'|), 
\end{equation} 
where the Green function is introduced 
\begin{equation} 
\label{green} 
G(r)=-{\rm P.P.}\int\frac{\d{\bf k}}{(2\pi)^D}\frac{e^{i{\bf k}\cdot{\bf r}}}{2(T_k-\mu)}. 
\end{equation} 
In the dilute limit, when the average distance between particles is
much less than the coherence length, the wave function
$\varphi(r)/n_0$, describing the behaviour of two particles in the
condensate, should be proportional~\cite{cherny,cherny5} to the
$s$-wave function $\varphi^{(0)}(r)$, which corresponds to relative
motion of two particles with zero momentum and obeys the two-body
Schr\"odinger equation in the center-of-mass system
\begin{equation} 
-(\hslash^2/m)\nabla^2\varphi^{(0)}(r)+V(r)\varphi^{(0)}(r)=0. 
\label{twobody} 
\end{equation} 

In the 3D case, the coefficient of proportionality is equal to unity~\cite{bog47} in the leading order with respect to the
density, provided the  following boundary conditions are imposed: first $|\varphi_{\rm 3D}^{(0)}(r)|< \infty$ at $r=0$
and second, $\varphi_{\rm 3D}^{(0)}(r)\to  1-a/r$ for $r\to\infty$. In the developed formalism, this can be easily
inferred from the obtained equation~(\ref{phir}).  Indeed, direct integration in Eq.~(\ref{green}) gives $G_{\rm
3D}(r)=-m\cos(\sqrt{2}r/\xi) /(4\pi\hslash^2r)$, and, hence, $G_{\rm 3D}(r)\simeq -m/(4\pi\hslash^2r)$ when $r\lesssim
\xi$. Thus we have $\varphi(r)\simeq n_0\varphi_{\rm 3D}^{(0)}(r)$ within this region, and integration of Eq.~(\ref{twobody})
yields  $U(0)=4\pi\hslash^2a/m$. For the dilute gas we have also $n_0\simeq n$, and Eq.~(\ref{muhom}) leads to the
familiar expression  for the chemical potential $\mu\simeq 4\pi\hslash^2na/m$. 

In the 2D case, the low-energy behaviour of the 2D  Green's function~(\ref{green}) is easily calculated: $G_{\rm 2D}(r)
\simeq m/(2\pi \hslash^2) \ln[e^\gamma r/(\sqrt{2}\xi)]$ when  $r\lesssim \xi$. Then it is not difficult to see from
Eq.~(\ref{phir})  that, first, $\varphi(r)/n_0$ obeys the 2D Schr\"odinger equation
(\ref{twobody}), and, second, its asymptotics for $r\to\infty$  is 
\begin{equation} 
\varphi(r)/n_0\to 1+\ln[e^\gamma r/(\sqrt{2}\xi)]mU(0)/(2\pi\hslash^{2}). 
\label{bound1} 
\end{equation} 
Hence, due to linearity of Eq.~(\ref{twobody}), the solution for $\varphi(r)$ should be proportional to
the wavefunction $\varphi^{(0)}_{\rm 2D}(r)$ that obeys the 2D Schr\"odinger equation~(\ref{twobody}) with the 
following boundary conditions: first $|\varphi_{\rm 2D}^{(0)}(r)|< \infty$ at $r=0$ second, 
$\varphi^{(0)}_{\rm 2D}(r)\to \ln(r/a_{\rm 2D})$ for $r\to\infty$. The latter equation can be considered as the definition 
of the 2D scattering length~\cite{lieb1}. Note that in the case of hard disks, $a_{\rm 2D}$ coincides with
the radius of the disks. It is convenient to introduce the dimensionless
parameter $u$ by the relation $U(0)=4\pi\hslash^{2}u/m$, such that $u$ is the dimensionless scattering amplitude
for two bosons in a medium of other bosons. By comparing the asymptotics~(\ref{bound1}) with that of $\varphi^{(0)}_{\rm
2D}(r)$, we derive 
\begin{eqnarray} 
\varphi(r)&=&2un_0\varphi^{(0)}_{\rm 2D}(r), 
\nonumber\\ 
-\ln (a_{\rm 2D}/\xi)&=&1/(2u)+\ln(e^\gamma/\sqrt{2}). 
\label{equ} 
\end{eqnarray} 
With the help of Eq.~(\ref{muhom}) and the definition of $\xi$ (see above), the relation~(\ref{equ}) becomes a 
self-consistent equation for $u$ 
\begin{equation} 
1/u+\ln u=-\ln(n_{\rm 2D}a_{\rm 2D}^{2}2\pi)-2\gamma, 
\label{udelta} 
\end{equation} 
where we neglect the condensate depletion in the leading order, putting $n_0\simeq n_{\rm 2D}$. By means of the latter 
approximation, the expression~(\ref{muhom}) takes the form 
\begin{equation} 
\mu={4\pi\hslash^{2}n_{\rm 2D}u}/{m}. 
\label{muhom2d} 
\end{equation} 
Thus, the 2D chemical potential is given by Eqs.~(\ref{udelta}) and  (\ref{muhom2d}), which lead to the density
expansion 
\begin{eqnarray} 
\mu&=&\frac{4\pi\hslash^{2}n_{\rm 2D}}{m}\Bigg(-\frac{1}{\ln(n_{\rm 2D}a_{\rm 2D}^{2})} 
+\frac{1}{\ln^2(n_{\rm 2D}a_{\rm 2D}^{2})} 
\nonumber\\ 
&&\times\ln\bigg[-\frac{1}{\ln(n_{\rm 2D}a_{\rm 2D}^{2})}\bigg] +\ldots 
\Bigg). 
\label{mu2dexp} 
\end{eqnarray} 
Equations~(\ref{udelta}) and (\ref{muhom2d}) are in agreement with
the results of Refs.~\cite{popov,kolom1,ovchin} and with the more
accurate scheme of Ref.~\cite{cherny4}, which yields the correction for the
chemical potential
\begin{equation}
\mu=(4\pi\hslash^{2}n_{\rm 2D}/m)(u+u^{2}+\cdots).
\label{muhom2dcorr}
\end{equation}
Here, $u$ is given by the more exact relation
\begin{equation}
1/u+\ln u = -\ln(n_{\rm 2D} a_{\rm 2D}^{2}\pi)-2\gamma,
\label{uphi2}
\end{equation}
where $\gamma=0.5772\ldots$ is the Euler constant. By means of this
relation, one can rewrite Eq.~(\ref{muhom2dcorr}) in terms of the gas
parameter $n_{\rm 2D}a_{\rm 2D}^{2}$ and obtain three more terms in
the expansion~(\ref{mu2dexp}). Note that Eq.~(\ref{uphi2}) differs
from Eq.~(\ref{udelta}) by a numerical factor under the logarithm,
which is essential only for obtaining these additional terms but not
the terms given by relation~(\ref{mu2dexp}).

\subsection{The inhomogeneous case} 
\label{sec:nonhom}
 
\subsubsection{The Gross-Pitaevskii regime}
\label{sec:gpregime} 

First of all, we should verify that the equations obtained in
Sec.~\ref{sec:gpgen} lead to the standard GP scheme in the case
$R_{\rm e}\ll \xi \ll l$, where $l$ is the characteristic length of an
isotropic trap. In this regime, one can expect that the pair wave
function $\varphi({\bf r}_1,{\bf r}_2)$ is very close to that obtained
in the homogeneous case, with the difference that the density is
spatially dependent now. So, we put by definition $\varphi({\bf
r}_1,{\bf r}_2) =\phi({\bf r}_1)\phi({\bf r}_2) \widetilde{\varphi}
({\bf r}_1,{\bf r}_2)$ and $\psi({\bf r}_1,{\bf r}_2) =\phi({\bf
r}_1)\phi({\bf r}_2)\widetilde{\psi}({\bf r}_1,{\bf r}_2)$, and,
hence, $\widetilde{\varphi}({\bf r}_1,{\bf r}_2)
=1+\widetilde{\psi}({\bf r}_1,{\bf r}_2)$ by
Eq.~(\ref{phidef}). Substituting those expressions into
Eqs.~(\ref{phieq}) and (\ref{psieq}) yields
\begin{eqnarray}
&&\bigg[-\frac{\hslash^2}{2m}\nabla_1^2-\mu -E_0 + V_{\rm ext}({\bf r}_1)\bigg]\phi({\bf r}_1)
+\phi({\bf r}_1)|\phi({\bf r}_1)|^{2}
\nonumber\\
&&\quad\quad\quad\times\int \d {\bf r}_2\, V(|{\bf r}_1-{\bf r}_2|)\widetilde{\varphi}({\bf r}_1,{\bf r}_2)=0,
\label{phieqsub}\\ 
&&-\frac{\hslash^2}{2m}(\nabla_1^2+\nabla_2^2)\widetilde{\psi}({\bf r}_1,{\bf r}_2)
+ V({\bf r}_1-{\bf r}_2)\widetilde{\varphi}({\bf r}_1,{\bf r}_2)
\nonumber \\
&&\quad\quad\quad=[f({\bf r}_1)+f({\bf r}_2)]\widetilde{\psi}({\bf r}_1,{\bf r}_2),
\label{psieqsub}
\end{eqnarray}
where we use the condition $R_{\rm e}\ll\xi$ in the first equation and introduce the notation 
\begin{align} \label{eqn:f}
f({\bf r})=\int\d {\bf r}'\,|\phi({\bf r}')|^{2}V({\bf r}-{\bf r}')\widetilde{\varphi}({\bf r},{\bf r}')
+\frac{\hslash^2}{m}\frac{\nabla_{\bf r}\phi({\bf r})}{\phi({\bf r})}\cdot\nabla_{{\bf r}},
\end{align}
with the last term being a differential operator.
Since $\varphi({\bf r}_1,{\bf r}_2)\simeq\phi({\bf r}_1)\phi({\bf
r}_2)$ at the distances of order of the correlation length, we have
$\widetilde{\varphi}({\bf r}_1,{\bf r}_2)\simeq 1$ at these
distances. Consequently, the l.h.s. of Eq.~(\ref{psieqsub}) remains
finite when the density tends to zero, while the r.h.s. becomes
small. Indeed, the first term of Eq.~(\ref{eqn:f}) is of order of
$\hslash^2an/m$. The second term is
less than $\hslash^2/(m\xi^{2})$ because the characteristic scale of the
order parameter cannot be smaller than $\xi$ in the case $\xi\ll l$
and the same applies to $\widetilde{\psi}$. Hence, in the leading
order we can completely neglect the r.h.s.\ of Eq.~(\ref{psieqsub}),
which leads to the standard Schr\"odinger equation (\ref{twobody}) for
$\widetilde{\varphi}$. Thus, we come to the approximation
\begin{equation}
\varphi({\bf r}_1,{\bf r}_2) \simeq \phi({\bf r}_1) \phi({\bf r}_2) \varphi_{\rm 3D}^{(0)}(r).
\label{gpphi}
\end{equation}
Using the well-known relation for the 3D scattering length
\begin{equation}
4\pi\hslash^2a/m=\int\d^{3}r\,V(r)\varphi_{\rm 3D}^{(0)}(r),
\label{iden}
\end{equation}
we can rewrite Eq.~(\ref{phieqsub}) in the standard GP form
with the coupling constant $g=4\pi\hslash^2a/m$. Note that,
nevertheless, the equilibrium value of the energy (\ref{efunc})
differs from that of the GP value (\ref{gpfun}) by the terms arising
from the condensate depletion because the second term in the r.h.s. of
Eq.~(\ref{onebody}) is not equal to zero. We will discuss these
corrections to the energy in Secs.~\ref{sec:kinpot} and \ref{sec:vir}.

\subsubsection{2D regime}
\label{sec:2dregime}

Here we consider the Bose gas confined only in $z$-direction by the
trapping potential $V_{\rm ext}=V_{\rm ext}(z)$. The system
is homogeneous in the $x$-$y$ plane and assumed to be infinitely
large. Physically this means that the $x$-$y$ size of the system is much
larger than the characteristic
radius of the trapping potential $l_{z} \equiv
\sqrt{\hslash/(m\omega_z)}$. The order parameter $\phi$ now becomes 
independent of $x$ and $y$, and the two-body function depends on the
relative distance $\rho=|\bm{\rho}_1-\bm{\rho}_2|$ between points
$\bm{\rho}_1=(x_1,y_1)$ and $\bm{\rho}_2=(x_2,y_2)$, so $\varphi({\bf
r}_1,{\bf r}_2)=\varphi(z_1,z_2,\rho)$. The 2D regime is provided by
the condition $l_z \ll\xi$.
Moreover, the condition $R_{\rm e}\ll\xi$ is fulfilled in most
experiments. As was discussed in Sec.~\ref{sec:intro} the density
profile is then governed by the ground state solution $\phi_0(z)$ of the
one-particle Schr\"odinger equation
\[
\big[-\hslash^2\nabla^2/(2m)-E_0 + V_{\rm ext}(z)\big]\phi_0(z)=0,
\]
because the second term in Eq.~(\ref{phieq}) can be treated as a small
correction. Thus, we can put in the leading order
$\phi(z)\simeq\sqrt{n_{\rm 2D}}\phi_0(z)$; $\phi_0(z)$ is normalized to unity. By analogy with
standard perturbation theory, the chemical potential, as the first
correction to $E_0$, can be found with the unperturbed eigenfunction
$\phi_0$. So, multiplying Eq.~(\ref{phieq}) by $\phi_0(z_1)$ and
integrating by $z_1$ yield
\begin{equation}
\mu=n_{\rm 2D}\int \d\bm{\rho}\, \widetilde{U}(\rho),
\label{mu2d}
\end{equation}
where by definition
\begin{eqnarray}
\widetilde{U}(\rho)&\equiv&\int \d z_1\d z_2\,
V(\sqrt{\rho^{2}+(z_1-z_2)^{2}}) \nonumber\\ 
&&\quad\times\varphi(z_1,z_2,\rho)\phi_0(z_1)\phi_0(z_2)/n_{\rm 2D}.
\label{deftilu}
\end{eqnarray}
In the same manner, one can multiply Eq.~(\ref{psieq}) by $\phi_0(z_1)\phi_0(z_2)$ and carry out the integration by $z_1$
and $z_2$, which results in the equation
\begin{equation}
2\big[\hslash^{2}\nabla_{\rho}^{2}/(2m)+\mu\big]\widetilde{\psi}(\rho)=\widetilde{U}(\rho)
\label{tilpsiu}
\end{equation}
for the function
\begin{equation}
\widetilde{\psi}(\rho)=\int \d z_1\d z_2\, \psi(z_1,z_2,\rho)\phi_0(z_1)\phi_0(z_2)/n_{\rm 2D}.
\label{tilpsi}
\end{equation}
Thus, we arrive at the system of equations (\ref{mu2d}) and (\ref{tilpsiu}), which coincides with that of (\ref{muhom}) and
(\ref{psik}) in homogeneous case if we put $U(k)= \int\d\bm{\rho}\, \widetilde{U}(\rho)e^{-i{\bf k} \cdot \bm{\rho}}$ and
perform the Fourier transformation of Eq.~(\ref{tilpsiu}). By the same method as in Sec.~\ref{sec:hom}, we obtain the
asymptotics for sufficiently large $\rho$ [physically, for $R_{\rm e}\ll\rho\ll\xi$, when only the first term  dominates in
Eq.~(\ref{tilpsiu})] 
\begin{equation}
\widetilde{\varphi}(\rho)\simeq 1+\ln[e^\gamma \rho/(\sqrt{2}\xi)]m\mu/(2\pi\hslash^{2}n_{\rm 2D}),
\label{tilphiasymp}
\end{equation}
where by definition
\begin{equation}
\widetilde{\varphi}(\rho)=\int \d z_1\d z_2\, \varphi(z_1,z_2,\rho)\phi_0(z_1)\phi_0(z_2)/n_{\rm 2D}
=1+\widetilde{\psi}(\rho).
\label{tilphi}
\end{equation}
The latter relation is due to Eqs.~(\ref{phidef}) and (\ref{tilpsi}).

In order to obtain the chemical potential in terms of the  3D
scattering length $a$ and the length $l_{z}$ of the trapping potential,
we use the following approximation~\cite{kagan}
\begin{equation}
\varphi(z_1,z_2,\rho)=C\varphi^{(0)}_{\rm 3D}(r)n_{\rm 2D}\phi_0(z_1)\phi_0(z_2)
\label{phiapp}
\end{equation}
in the region $r\ll l_{z}\ll\xi$, where  $r=\sqrt{\rho^{2}+(z_1-z_2)^{2}}$, and $\varphi^{(0)}_{\rm 3D}(r)$
denotes the 3D solution of the Schr\"odinger equation (\ref{twobody}) with asymptotics for $r\gg R_{\rm e}$
\begin{equation}
\varphi^{(0)}_{\rm 3D}(r)\simeq 1-a/r.
\label{phi3dasymp}
\end{equation}
Here the crucial point is that the constant $C\not=1$, which
determines the 2D behaviour of the system. If we substitute
Eq.~(\ref{phiapp}) into Eq.~(\ref{tilphi}) and take the integral, we
arrive at a new expression for $\widetilde{\varphi}(\rho)$. This
should be expanded with respect to the dimensionless variable $\rho/l_{z}$
and compared with Eq.~(\ref{tilphiasymp}). Since the main contribution
in that integral comes from the asymptotics (\ref{phi3dasymp}), one
can use it instead of the function $\varphi^{(0)}_{\rm 3D}(r)$
itself. By performing this procedure for the harmonic trapping
potential with $\phi_0(z)=\exp[-z^2/(2l_{z}^2)]/\sqrt{l_{z}\sqrt{\pi}}$, we
have
\[
\widetilde{\varphi}(\rho)\simeq C+\frac{2Ca}{l_{z}\sqrt{2\pi}}\ln[e^{\gamma/2}\rho/(2\sqrt{2}l_{z})].
\]
Comparing this relation with Eq.~(\ref{tilphiasymp}) yields
\begin{equation}
C=\sqrt{2\pi}\,l_{z}u/a,
\label{cpar}
\end{equation}
and the chemical potential is given by Eq.~(\ref{muhom2d}) with the dimensionless parameter $u$ obeying the equation
\begin{equation}
1/u+\ln u = \sqrt{2\pi}\,l_{z}/a-\gamma-\ln(16\pi n_{\rm 2D}l_{z}^{2}).
\label{u2dquasi}
\end{equation}
This result for $\mu$ is well consistent with relations (\ref{mu2d})
and (\ref{deftilu}). Indeed, substitution of Eq.~(\ref{phiapp}) with
constant (\ref{cpar}) into Eq.~(\ref{deftilu}) leads to
Eq.~(\ref{muhom2d}) provided that the relation (\ref{iden}) is
employed in conjunction with the approximation
$\exp[-(z_1-z_2)^{2}/(2l_{z}^{2})]\simeq1$ due to the integration with the
short-range potential with $R_{\rm e}\ll l_{z}$. In the leading order at
small 2D densities, expressions (\ref{muhom2d}) and (\ref{u2dquasi})
result in
\begin{equation}
\mu=\frac{4\pi\hslash^{2}n_{\rm 2D}}{m}\frac{1}{\sqrt{2\pi}\,l_{z}/a-\gamma-\ln(16\pi
n_{\rm 2D}l_{z}^{2})} . 
\label{muourapp}
\end{equation}
This differs from the result~\cite{shlyap2D} of Petrov, Holzmann, and
Shlyapnikov only by the additional numerical term
$-\gamma-\ln2=-1.2703\ldots$ in the denominator.
We note that the healing length in two dimensions takes the form $\xi=1/\sqrt{4\pi n_{\rm 2D}u}$, which differs from
that in three dimensions $\xi=1/\sqrt{4\pi na}$. Due to the criterion $1/\sqrt{n_{\rm 2D}}\ll\xi$, the obtained results
relate to sufficiently small densities, for which $u\ll 1$.

\subsubsection{1D regime}
\label{sec:1dregime}

Contrary to the 3D and 2D non-ideal Bose gases, there is no
Bose-Einstein condensate in one dimension~\cite{popov,hoh} in the
thermodynamic limit, because the long-wave fluctuations of the phase
break the off-diagonal long-range order.  Nevertheless, one can speak
about the quasi-condensate~\cite{shlyap1D} if a size of the of the 1D
system is sufficiently small. Indeed, at zero temperature the phase
fluctuations are suppressed if $\ln(L_{z}/\xi)\ll
n_{\rm 1D}\xi$~\cite{shlyap1D,ho}, which can be fulfilled only at finite number
of particles. Here $L_{z}$ stands for the size in
$z$-direction.

All calculations concerning the 1D quasi-condensate in the case
$R_{\rm e}\ll l_{\rho}\ll\xi$ can be done in complete analogy with the
2D inhomogeneous Bose gas considered in the previous subsection. The
gas is strongly confined in the $x$-$y$ plane by the harmonic trapping
potential in $V_{\rm ext} =m\omega_\rho^2 \rho^2/2$ with the length
$l_{\rho}=\sqrt{\hslash/(m\omega_\rho)}$, and remains homogeneous in
$z$-direction. In the regime involved, we can put
$\phi(\rho)=\sqrt{n_{\rm 1D}}\phi_0(\rho)$,
$\phi_0(\rho)=\exp[-\rho^2/(2l_{\rho}^2)]/l_{\rho}\sqrt{\pi}$ is the ground state
solution of the one-particle Schr\"odinger equation with the energy
$E_{0}=\hslash\omega_{\rho}$. Reasoning by analogy with
Sec.~\ref{sec:2dregime}, we obtain
\begin{equation}
\mu=n_{\rm 1D}\int \d z\, \widetilde{U}(z),
\label{mu1d}
\end{equation}
where we introduce the even function $\widetilde{U}(z)=\widetilde{U}(-z)$
\begin{eqnarray}
\widetilde{U}(z)&=&\int \d\bm{\rho}_1\d\bm{\rho}_2\, V(\sqrt{(\bm{\rho}_1-\bm{\rho}_2)^{2}+z^{2}}) \nonumber\\
&&\quad\times\varphi(\bm{\rho}_1,\bm{\rho}_2,z)\phi_0(\rho_1)\phi_0(\rho_2)/n_{\rm 1D}.
\nonumber
\end{eqnarray}
The 1D analogue of Eq.~(\ref{tilpsiu}) is the equation 
\begin{equation}
2\bigg[\frac{\hslash^{2}}{2m}\frac{\d^{2}}{\d z^{2}}+\mu\bigg]\widetilde{\psi}(z)=\widetilde{U}(z)
\label{tilpsiu1}
\end{equation}
for the function
\[
\widetilde{\psi}(z)=\int \d \bm{\rho}_1\d \bm{\rho}_2\, \psi(\bm{\rho}_1,\bm{\rho}_2,z)
\phi_0(\rho_1)\phi_0(\rho_2)/n_{\rm 1D}.
\]
Equation (\ref{tilpsiu1}) can be rewritten in the Lippmann-Schwinger form at $\mu\to0$ (see discussion in
Sec.~\ref{sec:hom}) 
\begin{equation}
\widetilde{\varphi}(z)=1+(m/\hslash^{2})\int \d z'\,\widetilde{U}(z')|z-z'|/2
\label{lipshw1}
\end{equation}
for the function $\widetilde{\varphi}(z)$, defining as 
\begin{equation}
\widetilde{\varphi}(z)=\int \d \bm{\rho}_1\d \bm{\rho}_2\, \varphi(\bm{\rho}_1,\bm{\rho}_2,z)
\phi_0(\rho_1)\phi_0(\rho_2)/n_{\rm 1D}.
\label{tilphi1}
\end{equation}
Equations (\ref{mu1d}) and (\ref{lipshw1}) give the asymptotics for $R_{\rm e}\ll z\ll\xi$
\begin{equation}
\widetilde{\varphi}(z)\simeq 1+m\mu|z|/(2n_{\rm 1D}\hslash^{2}).
\label{asymp1}
\end{equation}
On the other hand, in the region $r\ll l_{\rho}\ll\xi$ we can use the analogue of Eq.~(\ref{phiapp})
\begin{equation}
\varphi(\bm{\rho}_1,\bm{\rho}_2,z)=C\varphi^{(0)}_{\rm 3D}(r)n_{\rm 1D}\phi_0(\rho_1)\phi_0(\rho_2),
\label{phiapp1}
\end{equation}
which leads to the asymptotics after the integration in Eq.~(\ref{tilphi1})
\begin{equation}
\widetilde{\varphi}(z)\simeq C-C\big(\sqrt{\pi/2}-|z|/l\big)a/l_{\rho}.
\label{asymp11}
\end{equation}
Comparing Eqs.~(\ref{asymp1}) and (\ref{asymp11}) yields
\begin{eqnarray}
C&=&1/(1-\sqrt{\pi/2}\,a/l_{\rho}),
\label{cpar1}\\
\mu&=&\frac{2\hslash^{2}n_{\rm 1D}}{m}\frac{a}{l_{\rho}^{2}}\frac{1}{1-\sqrt{\pi/2}\,a/l_{\rho}},
\label{mu1d1}
\end{eqnarray}
which differs from Olshanii's result~\cite{olshanii} through the
numerical factors $\sqrt{\pi}=1.772\ldots$ in the denominator instead
of the constant $1.4603\ldots$ introduced by him. We note that in
the paper~\cite{olshanii}
$a_{\perp}=\sqrt{2\hslash/(m\omega_\rho)}=\sqrt{2}\,l_{\rho}$ in our
notation. One can see that the criteria of applicability of the
obtained results $l_{\rho}\ll\xi$ and $1/n_{\rm 1D}\ll\xi$ impose the following
restriction on 
the 1D density
\begin{equation}
\frac{a}{l_{\rho}^{2}}\ll n_{\rm 1D}\ll\frac{1}{a},
\label{restr1d}
\end{equation}
since $\xi\simeq l_{\rho}/\sqrt{2an_{\rm 1D}}$ in one dimension.

\section{The kinetic, interaction, and external field energy of the trapped Bose gas} 
\label{sec:kinpot} 

The simplest method for obtaining the values of the interaction
energy~(\ref{eint}), the kinetic energy (\ref{ekin}), and the energy
of interaction with an external field (\ref{eext}), is to apply the
variational theorem. The latter can be formulated in general as
follows. If a function $f(x)$ obeys the functional equation
\begin{equation}
\delta F[\{f(x)\},\lambda]/\delta f(x)=0
\label{fan}
\end{equation}
with the functional $F$ depending on the function $f(x)$ and the
parameter $\lambda$, then the solution of Eq.~(\ref{fan})
$f(x)=f_0(x,\lambda)$ is also dependent on $\lambda$. Nevertheless,
when calculating the derivative of the stationary value of the
functional with respect to $\lambda$, we can take into consideration
only the explicit dependence on this parameter
\begin{equation}
\d F[\{f_0(x,\lambda)\},\lambda]/\d \lambda=\partial F[\{f_0(x,\lambda)\},\lambda]/\partial \lambda.
\label{fander}
\end{equation}
This is obvious due to Eq.~(\ref{fan}).

The variational theorem (\ref{fander}) is still valid if the
functional contains two or more functions. In our case, the
functions can be associated with $\phi(x_1)$ and $\psi(x_1,x_2)$
involved in the energy functional~(\ref{efunc}).  Considering ${\cal
N}$ as the parameter of the variational theorem, we come to the
standard thermodynamic relation $\partial E/\partial {\cal
N}=\mu'=\mu+E_0$. One can rewrite this derivative in terms of the
energy per particle $\varepsilon=E/N$ and the density of particles
$\partial E/\partial {\cal N} =\partial(\varepsilon n)/\partial n$,
which gives the relation $\varepsilon=(1/n)\int_{0}^{n} \d
n'\,\mu(n')+E_0$. Then relations~(\ref{muhom2d}) and (\ref{u2dquasi})
lead to
\begin{equation} 
\varepsilon_{\rm 2D}\simeq{2\pi\hslash^{2}n_{\rm 2D}u}/{m}+\frac{\hslash^2}{2ml_{z}^{2}}
\label{e2d} 
\end{equation}
with $u$ given by Eq.~(\ref{u2dquasi}). In the same
manner, we obtain from Eq.~(\ref{mu1d1})~\cite{note2}
\begin{equation}
\varepsilon_{\rm 1D}=\frac{\hslash^{2}n_{1D}}{m}\frac{a}{l_{\rho}^{2}}\frac{1}{1-\sqrt{\pi/2}\,a/l_{\rho}}
 +\frac{\hslash^2}{ml_{\rho}^{2}}.
\label{e1d}
\end{equation}
Equations (\ref{e2d}) and
(\ref{e1d}) give us the equilibrium value of the energy (\ref{efunc})
per particle in the 2D and 1D cases, respectively. In order to
calculate the interaction energy with the help of the variational
theorem, one can replace $V\to\lambda V$ and differentiate
$\varepsilon$ with respect to $\lambda$ at $\lambda=1$. 
All we need to know is the derivative of the 3D scattering length, which reads~\cite{cherny1,mur} 
\begin{equation}
\lambda \frac{\partial a}{\partial \lambda}=
m\frac{\partial a}{\partial m}
=\frac{m}{4\pi\hslash^2}\int\d^{3}r\,[\varphi_{\rm 3D}^{(0)}(r)]^{2}\lambda V(r).
\label{varth}
\end{equation}
It is convenient to introduce one more characteristic length~\cite{cherny1}, the positive parameter $b$,
\begin{eqnarray}
b=a-\lambda\left.\frac{\partial a}{\partial \lambda}\right|_{\lambda=1}
=\frac{1}{4\pi}\int\d^{3}r\,\bigl|\nabla\varphi_{\rm 3D}^{(0)}(r)\bigr|^{2}.
\nonumber
\end{eqnarray} 
So, we have
\begin{eqnarray}
\varepsilon_{\rm 2Dint}&\simeq&\frac{2\pi\hslash^{2}n_{\rm 2D}}{m}u^{2}\frac{\sqrt{2\pi}\,l_z}{a}\bigg(1-\frac{b}{a}\bigg),
\label{eint2d}\\
\varepsilon_{\rm 1Dint}&\simeq&\frac{\hslash^{2}n_{1D}}{m}\frac{a}{l_{\rho}^{2}}\bigg(1-\frac{b}{a}\bigg),
\label{eint1d}
\end{eqnarray}
where we use the approximation $u^2/(1-u)\simeq u^2$ in
Eq.~(\ref{eint2d}) and restrict ourselves by the leading order in
Eq.~(\ref{eint1d}).

With the same method, replacing $V_{\rm ext}\to\lambda V_{\rm
ext}$ (which is equivalent to $l\to l/\sqrt[4]{\lambda}$) and
differentiating, we arrive at the external energy per particle
\begin{eqnarray}
\varepsilon_{\rm 2Dext}&\simeq&\frac{2\pi\hslash^{2}n_{\rm
2D}}{m}\frac{u^{2}}{4}\bigg(\frac{\sqrt{2\pi}\,l_z}{a}-2\bigg)+\frac{\hslash^2}{4ml_{z}^{2}},
\label{eext2d}\\
\varepsilon_{\rm 1Dext}&\simeq&\frac{\hslash^{2}n_{1D}}{2m}\frac{a}{l_{\rho}^{2}}
+\frac{\hslash^2}{2ml_{\rho}^{2}}.
\label{eext1d}
\end{eqnarray}
In the same manner, we have $\varepsilon_{\rm kin}=-m\partial\varepsilon/\partial m$, which leads to
\begin{eqnarray}
\varepsilon_{\rm 2Dkin}&\simeq&\frac{2\pi\hslash^{2}n_{\rm 2D}}{m}u 
-\frac{2\pi\hslash^{2}n_{\rm 2D}}{m}u^2\bigg[\frac{1}{4}\bigg(\frac{\sqrt{2\pi}\,l_z}{a}-2\bigg)\nonumber\\
&&+\frac{\sqrt{2\pi}\,l_z}{a}\bigg(1-\frac{b}{a}\bigg)\bigg]+\frac{\hslash^2}{4ml_{z}^{2}},
\label{ekin2d}\\
\varepsilon_{\rm 1Dkin}&\simeq&\frac{\hslash^{2}n_{1D}}{m}\frac{b}{l_{\rho}^{2}}
-\frac{\hslash^{2}n_{1D}}{2m}\frac{a}{l_{\rho}^{2}}+\frac{\hslash^2}{2ml_{\rho}^{2}}.
\label{ekin1d}
\end{eqnarray}
One can see that sum of the kinetic, external and interaction energies
equals to the total energy, as it should be.  Note that the developed
formalism allows us to calculate the interaction energy directly,
starting from the expression (\ref{eint1}) and using
Eq.~(\ref{iden}), since we have the analytic expressions (\ref{phiapp}),
(\ref{cpar}), (\ref{phiapp1}), and (\ref{cpar1}) for the short-range
behaviour of the anomalous average.

We note that the ratio $b/a$ need not be small. In particular, it is
of order of ten for the realistic interaction potentials of alkali
atoms~\cite{cherny3}. We stress that the term with the length $b$
appears in the mean interaction energy by virtue of the the
short-range two-body correlations at the distances of order of $a$ and
in the mean kinetic energy by sufficiently large momenta of order of
$p\gtrsim \hslash/a$ in the momentum distribution. In the static
structure factor, this region is rather difficult to be measured
experimentally.  The density expansion method gives the value of the
release energy that is defined as {\it sum} of the interaction and
kinetic energies
\begin{eqnarray}
\varepsilon_{\rm 2Drel}&\simeq&\frac{2\pi\hslash^{2}n_{\rm 2D}}{m}u 
-\frac{2\pi\hslash^{2}n_{\rm 2D}}{m}\frac{u^2}{4}\bigg(\frac{\sqrt{2\pi}\,l_z}{a}-2\bigg)\nonumber\\
&&+\frac{\hslash^2}{4ml_{z}^{2}},
\label{erel2d}\\
\varepsilon_{\rm 1Drel}&\simeq&\frac{\hslash^{2}n_{1D}}{2m}\frac{a}{2l_{\rho}^{2}}+\frac{\hslash^2}{2ml_{\rho}^{2}}.
\label{erel1d}
\end{eqnarray}
As one can see, the parameter
$b$ is canceled and not involved in the values of the release energy.
Let us compare the values of the release (\ref{erel2d}-\ref{erel1d}) and total energy (\ref{e2d}-\ref{e1d}). The 
energy of zero-point oscillation is involved in the release energy with the factor $1/2$, as it should be for 
the harmonic trap. The other terms would coincide in the standard GP approach, but we have obvious difference
due to accounting for the non-condensate contribution.
In principle, the obtained corrections should be measurable in experiments.

\section{Virial theorem} 
\label{sec:vir}
 
The virial theorem can be obtained immediately from the energy
functional (\ref{efunc}) if we consider its variation in vicinity of
the stationary state (ground state) with respect to the scaling
transformation of the ground state functions $\phi_0$ and $\psi_0$,
obeying the generalized GP equations (\ref{phieq}) and
(\ref{psieq}). Namely, we substitute into Eq.~(\ref{efunc}) the
functions $\phi({\bf r}_1) =\alpha^{3/2}\phi_0(\alpha {\bf r}_1)$ and
$\psi({\bf r}_1) =\alpha^{3}\psi_0(\alpha {\bf r}_1,\alpha {\bf
r}_1)$. Replacing the variables in the integrals ${\bf r}_1\to\alpha
{\bf r}_1$ and ${\bf r}_2\to\alpha {\bf r}_2$, we notice that, first,
the last term in the functional equals to zero for any $\alpha$, and,
second, the other terms can be written in terms of its stationary
values
\begin{equation}
E(\alpha)=\alpha^{2}E_{\rm kin}+E_{\rm ext}/\alpha^{2}+E_{\rm int}\big[V(r/\alpha)\big].
\label{ealph}
\end{equation} 
Since the variation of the functional should be zero for any small
variations of the functions, we have $\d E/\d \alpha=0$ at $\alpha=1$,
which leads to
\begin{equation}
2E_{\rm kin}-2E_{\rm ext}+E_{\rm int}[-rV'(r)]=0,
\label{virth}
\end{equation} 
where the terms are given by Eqs.~(\ref{eint1})-(\ref{ekin1}). The
value of the last term corresponds to the interaction energy with the
potential $-rV'(r)=-r\d V(r)/\d r$. In the case of the GP
approximation (\ref{gpphi}), one can simplify the last item in
Eq.~(\ref{virth}) by means of Eq.~(\ref{iden}) and relation~\cite{bog47}
\[
\frac{4\pi\hslash^2 a}{m}=-
\int_{0}^{\infty}\d r\, 4\pi r^2 [\varphi^{(0)}(r)]^{2}\bigg(2V(r)+r\frac{\d V(r)}{\d r}\bigg).
\]
The result takes a form
\begin{eqnarray}
E_{\rm int}&\simeq&\frac{1}{2}\int\d{\bf R}\,|\phi({\bf R})|^{4}
\int \d {\bf r}\,[-rV'(r)]\big[\varphi_{\rm 3D}^{(0)}(r)\big]^2
\nonumber\\
&=&\frac{2\pi\hslash^2}{m}(3a-2b)\int\d{\bf R}\,|\phi({\bf R})|^{4}.
\label{gpvirth}
\end{eqnarray}
If the potential is of the weak-coupling type~\cite{classif}, one can
neglect $b\ll a$ and arrive at the virial theorem obtained for the
$\delta$-function interaction potential~\cite{dalfovo}.

If the system is homogeneous in the $x$-$y$ plane (the 2D Bose gas of
Sec.~\ref{sec:2dregime}) or in the $z$ direction (the 1D Bose gas of
Sec.~\ref{sec:1dregime}), it can be considered as confined by infinite
walls in associated directions. Then one should be careful when
deriving the virial theorem from Eq.~(\ref{ealph}), as all its terms
relate to the density $n_{\rm 2D}/\alpha^2$ or $n_{\rm 1D}/\alpha$ for
the 2D or 1D Bose gas, respectively. For this reason, we come to
\begin{eqnarray}
2n_{\rm 2D}\frac{\partial\varepsilon_{\rm 2D}}{\partial n_{\rm 2D}}\!\!&=&\!\!2\varepsilon_{\rm 2D kin}\!
-2\varepsilon_{\rm 2D ext}\!+\varepsilon_{\rm 2D int}[-rV'(r)],
\label{virth2d}\\
n_{\rm 1D}\frac{\partial\varepsilon_{\rm 1D}}{\partial n_{\rm
1D}}\!\!&=&\!\!2\varepsilon_{\rm 1D kin}\!  -2\varepsilon_{\rm 1D
ext}\!+\varepsilon_{\rm 1D int}[-rV'(r)].
\label{virth1d}
\end{eqnarray} 
The interaction term in these equations can be easily calculated by
analogy with Eq.~(\ref{gpvirth}) but using Eqs.~(\ref{phiapp}) and
(\ref{phiapp1}), respectively. It is not difficult to be convinced
with the help of Eqs.~(\ref{e2d}) and (\ref{e1d}) that the virial
theorems (\ref{virth2d}) and (\ref{virth1d}) are fulfilled.

One can also find a relation between the chemical potential and the
various parts of the energy. Let us multiply Eq.~(\ref{phieq}) by
$\phi(x_1)$ and integrate over $x_1$, and multiply Eq.~(\ref{psieq})
by $\psi(x_1,x_2)$ and also integrate over $x_1$ and $x_2$. Summing
the obtained expressions yields
\begin{equation}
N\mu=E_{\rm kin1}+2E_{\rm kin2} +E_{\rm ext1}+ 2E_{\rm ext2}+2E_{\rm int},
\label{muenergy}
\end{equation}
Here, $E_{\rm ext1}$ and $E_{\rm kin1}$ are the condensate
contributions in the external and kinetic energies given by the last
terms in Eqs.~(\ref{eext1}) and (\ref{ekin1}), respectively, and
$E_{\rm ext2}$ and $E_{\rm kin2}$ are associated with the
non-condensate contributions, given by the residual parts of these
equations. One can easily see that the relation (\ref{muenergy}) is
fulfilled with $E_{\rm ext1}$ and $E_{\rm kin1}$ corresponding to the
last terms in Eqs.~(\ref{eext2d}) and (\ref{eext1d}), and
(\ref{ekin2d}) and (\ref{ekin1d}) for the 2D and 1D Bose gases,
respectively. One can notice that $E_{\rm kin2}$ could be negative for
the 1D Bose gas, if $b<a/2$ [see the first two terms in
Eq.~(\ref{ekin1d})].  Certainly, this is not a drawback of
Eqs.~(\ref{phieq}) and (\ref{psieq}) it is but due to the choice of
anzatz $\phi(\rho)=\sqrt{n_{\rm 1D}}\phi_0(\rho)$, which leads to
overestimation of the quasicondensate contribution $E_{\rm kin1}$ in
the 1D kinetic energy. Indeed, the Gaussian profile $n_{\rm
1D}|\phi_0(\rho)|^{2}$ relates to the {\it total} density of the 1D
gas $\langle{\hat\Psi}^{\dag}({\bf r}) {\hat\Psi}({\bf r})\rangle$ but not
to the ``quasicondensate component'' $|\phi(\rho)|^{2}$.  The latter is
difficult to define accurately in the 1D case, since there is no
eigenvalue of the one-body density matrix that is proportional to the
total number of particles. Nevertheless, we stress that the total
value of $E_{\rm 1Dkin}$ is positive, and the results (\ref{eint1d}),
(\ref{eext1d}), and (\ref{ekin1d}) look quite reasonable.

\section{Conclusions} 
\label{sec:dis} 
 
The main result of this paper are the generalized GP equations in the
time-dependent (\ref{phieqt}--\ref{psieqt}) and stationary form
(\ref{phieq}--\ref{psieq}), which allow us to determine the
interaction term self-consistently for interaction
potentials even containing a hard-core. The method,
which can be used for homogeneous, strongly inhomogeneous
quasi-low-dimensional, and cross-over regimes was derived within a
general HFB framework.

The HFB method is a mean-field approximation, which generally works
well only for weak-coupling potentials~\cite{classif}. In order to
extend the HFB scheme to hard-core potentials, the bare interaction
potential is usually replaced by a renormalized pseudopotential
$V(r)\to(4\pi\hslash^2/m)\delta^3({\bf r})$. However, such a
replacement leads to an
ultraviolet divergence and incorrect treatment of short-range
correlations of the particles. We have shown that the appropriate
renormalization can be obtained {\em within} the HFB scheme if, from
the two-body density matrix, only the anomalous correlation function
$\varphi(x_1,x_2) =\langle{\hat\Psi}(x_1) {\hat\Psi}(x_{2})\rangle$ is
retained. The anomalous correlation function can be interpreted as the
wavefunction of two bosons in the condensate. Its short-range
behaviour is described well in the proposed scheme at the cost of
loosing the correct description of the long-range behaviour.
However, long-range
correlations are not needed for deriving the non-linear term in the
generalized GP approach, which instead is determined by short-range
correlations.  Methods which can describe both the short- and
long-range correlations accurately were discussed in
Refs.~\cite{cherny4,cherny1,cherny2,leggett1}, but these methods
are appropriate only for the homogeneous Bose gas. The method proposed
in this paper was shown to work as well in inhomogeneous
situations. Cigar (quasi-1D) and pancake (quasi-2D) geometries were
considered as examples. Furthermore, it was shown that the
contribution of short-range correlations to the kinetic and release
energies of a tightly trapped gas can be calculated within this scheme
and that they are substantial. Interesting future applications of the
proposed method may include the modification of the nonlinearity in
quasi-1D waveguides \cite{Muryshev2002a,sinha04ep} and molecular Bose
condensates in optical lattices.

\section*{Acknowledgements}
The authors are grateful to O.~S\o{}rensen and
A.A.~Shanenko for interesting discussions. This work was supported in
part by the RFBR grant 01-02-17650.

\appendix 

\section{Two-body wave functions in the Hartree-Fock-Bogoliubov approximation}
\label{sec:pwf}

In general, the two-body density matrix can be expanded in a complete
set of its eigenfunctions
\begin{eqnarray}
\langle{\hat\Psi}^{\dagger}(x_1){\hat\Psi}^{\dagger}(x_{2}) 
{\hat\Psi}(x'_{2}){\hat\Psi}(x'_1)\rangle&=&
\sum_{\nu,\mu}N_{\nu,\mu}\varphi^*_{\nu,\mu}(x_1,x_2)
\nonumber\\
&&\times\varphi_{\nu,\mu}(x'_1,x'_2).
\label{rho2exp}
\end{eqnarray}
The eigenfunctions can be called two-body or pair wave functions. If
they are normalized to unity, it follows from Eq.~(\ref{rho2exp}) that
$\int\d x_1 \d x_2\,\langle{\hat\Psi}^{\dagger}(x_1)
{\hat\Psi}^{\dagger}(x_{2}) {\hat\Psi}(x_{2}) {\hat\Psi}(x_1)\rangle=
N(N-1)=\sum_{\nu,\mu}N_{\nu,\mu}$, i.e., the sum of all $N_{\nu,\mu}$
is the total number of pairs.  Therefore, the non-negative quantity
$N_{\nu,\mu}$ can be interpreted as the mean number of the pairs in
the state $(\nu,\mu)$, any pair being doubly taken.
 
Let us consider the homogeneous spinless Bose gas in the
HFB approximation \cite{griffin,hohmartin}. Within that approximation, the
two-body wave functions can be easily calculated~\cite{cherny}. The
statistical average of any product of quantum operators
$\hat{\vartheta}$ and $\hat{\vartheta}^\dag$ can be calculated with
the Wick-Bloch-De Dominicis theorem~\cite{bloch}, since the
Hamiltonian is approximated by a {\it quadratic form} of the Bose
operators ${\hat\alpha}_{{\bf p}}^{\dagger}$ and ${\hat\alpha}_{{\bf
p}}$ connected with initial operators ${\hat a}_{{\bf p}}^{\dagger}$
and ${\hat a}_{{\bf p}}$ by the canonical Bogoliubov transformations
(see Appendix~\ref{sec:hfrel}).  Extracting the $c$-number part
$\hat{\Psi} =\sqrt{n_0} +\hat{\vartheta}$ and $\hat{\Psi}^\dag
=\sqrt{n_0} +\hat{\vartheta}^\dag$ and using that theorem, one can
rewrite the four-boson average in the form
\[
\bigg\langle{\hat\Psi}^{\dag}\Big({\bf R}+\frac{\bf r}{2}\Big){\hat\Psi}^{\dag}\Big({\bf R}-\frac{\bf r}{2}\Big) 
{\hat\Psi}\Big({\bf R}'-\frac{{\bf r}'}{2}\Big){\hat\Psi}\Big({\bf R}'+\frac{{\bf r}'}{2}\Big)\bigg\rangle
\]
\vspace*{-.5cm}\begin{eqnarray}
&=&\!n_{0}^{2}\widetilde{\varphi}^{*}(r)\widetilde{\varphi}(r') +\int \d^{3}p\,\d^{3}q\,
\biggl[2n_{0}\delta({\bf q}/2-{\bf p})\frac{n(q)}{(2\pi)^{3}} \nonumber\\
&&\!+\frac{n({\bf q}/2+{\bf p})}{(2\pi)^{3}}
\frac{n({\bf q}/2-{\bf p})}{(2\pi)^{3}}\biggr]\sqrt{2}\cos({\bf p}\cdot\r{})\sqrt{2}\cos({\bf p}\cdot\rp{})\nonumber\\
&&\times\exp[i{{\bf q}}\cdot({\bf R}^{\prime }-{\bf R})],
\label{rho2exphf}
\end{eqnarray}
where we put by definition $\widetilde{\varphi}(r)
=1+\langle\hat{\vartheta}({\bf R}+{\bf r}/{2}) \hat{\vartheta}({\bf
R}-{\bf r}/{2})\rangle/n_0$. Because the
expansion~(\ref{rho2exphf}) is written in the thermodynamic limit, the
sum in Eq.~(\ref{rho2exp}) becomes an integral. The Bose-Einstein
condensate manifests itself in presence of $\delta$-functions in this
integral (note that the first term in the r.h.s. can be included in
the integral with the help of the $\delta$-functions). By comparing
the representation (\ref{rho2exp}) with that of (\ref{rho2exphf}), one
can conclude the following:\\ 
({\it i}) The quantum numbers of the
pair wave functions are the relative momentum $\nu={\bf p}$ and the
center-of-mass (total) momentum $\mu={\bf q}$ of two particles; all
these functions belong to continuous spectrum and thus describe the
scattering of two bosons in the medium of the other bosons.\\ 
({\it ii})
The maximum eigenvalue $N_{0}(N_{0}-1)\simeq N_{0}^{2}$ with ${\bf
p}={\bf q}=0$ corresponds to the state of two particles in the
condensate; its normalized eigenfunction $\widetilde{\varphi}(r)/V$
can be interpreted as a pair wave function of the
condensate-condensate type. Thus, the anomalous average
$\langle\hat{\vartheta}({\bf r})\hat{\vartheta}(0)\rangle$ can be
associated with the scattering part of the two-body wave function of
the bosons in the condensate~\cite{cherny}; in particular, it is
responsible for the short-range spatial correlations of two bosons in
the Bose-Einstein condensate.\\ 
({\it iii}) The other macroscopic
eigenvalues $2N_{0}n_{q}$ with ${\bf q}=\pm 2{\bf p}$ correspond to
the two-body states with one particle in the condensate and another
one beyond the condensate; its eigenfunctions $\sqrt{2}\cos({\bf
q}\cdot{\bf r}/2) \exp[i{\bf q}\cdot{\bf R}]/V$ are of the
condensate-noncondensate type~\cite{note3b}. The
residuary non-macroscopic  
eigenvalues $n({\bf q}/2+{\bf p}) n({\bf q}/2-{\bf p})$ are related to
the noncondensate-noncondensate pairs with the two-body wave
functions $\sqrt{2}\cos({{\bf p}}\cdot{\bf r}) \exp[i{\bf q}\cdot{\bf
R}]/V$.

Note that the wave function of the condensate-condensate type is not
reduced to a product of two one-body wave functions in the condensate,
which equal to $1/\sqrt{V}$ for the homogeneous Bose gas. This is
obvious, as particles in the Bose-Einstein condensate interact with
each other and with the other particles beyond the condensate. Another
important point is that all the other two-body wave functions are
symmetrized plane waves (consistent with the Born approximation) in
the framework of the HFB method. This is evidently a disadvantage of
the HFB scheme. As a consequence, we always arrive at divergences for a
hard-core potential when evaluating the contribution of the
condensate-noncondensate and noncondensate-noncondensate wave
functions in the interaction energy (\ref{eint}). At the same time,
the contribution of the condensate-condensate ``channel'' should be
finite in the interaction energy provided the anomalous averages are
calculated in a self-consistent manner. The generalization of the
expansion~(\ref{rho2exphf}) beyond the HFB approach and more detailed
discussions can be found in Ref.~\cite{cherny}. The pair wave function method of Ref.~\cite{cherny} was generalized to
the inhomogeneous systems in Ref.~\cite{naidon}.

\section{Relation between the normal and anomalous two-boson averages} 
\label{sec:hfrel} 
 
Let us establish a relation between the normal
$\langle{\hat\vartheta}^{\dag}(x_1){\hat\vartheta}(x_2)\rangle$ and the
anomalous average $\langle{\hat\vartheta}(x_1){\hat\vartheta}(x_2)\rangle$
for the vacuum state, which describes the behaviour of
the $N$-body system at zero temperature, in the framework of the
Hartree-Fock-Bogoliubov method. We remember that the vacuum state
$|0\rangle$ is defined as $\alpha_\nu|0\rangle=0$ for any $\nu\not=0$,
here the quasiparticle creation and destruction operators
$\hat{\alpha}^\dag_\nu$ and
$\hat{\alpha}_\nu$ can be introduced through
the Bogoliubov transformation ($f\not=0$)
\begin{eqnarray} 
\hat{a}_f&=&\sumpr_\nu\,(u_{f\nu}\hat{\alpha}_\nu+v_{f\nu}\hat{\alpha}^\dag_\nu),
\label{bogtrans_a} \\ 
\hat{a}^\dag_f&=&\sumpr_\nu\,(u^*_{f\nu}\hat{\alpha}^\dag_\nu+v^*_{f\nu}\hat{\alpha}_\nu), 
\label{bogtrans_b} 
\end{eqnarray} 
where $f$ and $\nu$ denote discrete (multi-)indices. The sum
$\sumpr_\nu$ means $\sum_{\nu\not=0}$ and the Bose-operators
$\hat{a}^\dag_f$ and $\hat{a}_f$ create and destruct a particle in the
eigenstate $\phi_{f}(x)$ of the one-body matrix
$\langle{\hat\Psi}^{\dag}(x'){\hat\Psi}(x)\rangle$
\[ 
\int \d x'\, \langle{\hat\Psi}^{\dag}(x'){\hat\Psi}(x)\rangle \phi_{f}(x') 
= n_{f} \phi_{f}(x), 
\] 
normalized as $\int d x\,|\phi_{f}(x)|^{2}=1$. Note that the set of
eigenfunctions including the normalized condensate function
$\phi_0(x)=\langle\hat{\Psi}(x)\rangle/\sqrt{N_0}$ with $N_0=n_{f=0}$
is complete and orthogonal
\begin{eqnarray} 
\sum_f\phi^*_f(x)\phi_f(x')=\delta(x-x'),
\label{compl_a}\\ 
\int \d x\,\phi_f^*(x)\phi_{f'}(x)=\delta(f-f'), 
\label{compl_b} 
\end{eqnarray} 
where we define the ``discrete" $\delta$-function as 
\[ 
\delta(f)=\left\{\begin{array}{ll} 
1, & f=0,\\ 
0, & f\not=0. 
\end{array}\right. 
\] 
From the Bose commutation relations
$[\hat{a}_f,\hat{a}\dag_{f'}]=\delta(f-f')$ and
$[\hat{\alpha}_f,\hat{\alpha}\dag_{f'}]=\delta(f-f')$ and
Eqs.~(\ref{bogtrans_a}-\ref{bogtrans_b}) we obtain at $f,f'\not=0$
\begin{eqnarray} 
\sumpr_\nu\,(u_{f\nu}u^*_{f'\nu}- v_{f\nu}v^*_{f'\nu})&=&\delta(f-f'), \label{uv_a}\\ 
\sumpr_\nu\,(u_{f\nu}v_{f'\nu}- v_{f\nu}u_{f'\nu})&=&0. 
\label{uv_b} 
\end{eqnarray} 
By using the definition of the quasiparticle vacuum state and
Eqs.~(\ref{bogtrans_a}-\ref{bogtrans_b}), we can calculate the averages
\begin{eqnarray} 
F(f,f')&=&\langle\hat{a}^\dag_f\hat{a}_{f'}\rangle=\sumpr_\nu\,v^*_{f\nu}v_{f'\nu},
\label{fphi_a}\\ 
\Phi(f,f')&=&\langle\hat{a}_f\hat{a}_{f'}\rangle=\sumpr_\nu\,u_{f\nu}v_{f'\nu}. 
\label{fphi_b} 
\end{eqnarray} 
Our purpose is to find the relation between the normal $F(f,f')$ and
the anomalous $\Phi(f,f')$ averages for that state.  In order to
simplify our calculations, we rewrite Eqs.~(\ref{bogtrans_a}-\ref{bogtrans_b}) in the
matrix notations
\begin{equation}\label{bogtransmat} 
\begin{pmatrix} 
\hat{a}\\ 
\hat{a}^\dag 
\end{pmatrix}=X 
\begin{pmatrix} 
\hat{\alpha}\\ 
\hat{\alpha}^\dag 
\end{pmatrix},\ 
X=\begin{pmatrix} 
U  & V\\ 
V^* & U^* 
\end{pmatrix}. 
\end{equation} 
Here the matrix $X$ is composed of the matrix $(U)_{ij}=u_{ij}$ and
$(V)_{ij}=v_{ij}$. The columns contain the operators $\hat{a}_f$
and $\hat{a}^\dag_f$, and $\hat{\alpha}_\nu$ and
$\hat{\alpha}^\dag_\nu$, respectively. We use the standard notations
for the complex conjugate $(V^*)_{ij}=v^*_{ij}$, transposed
$(V^\tran)_{ij}=v_{ji}$, and Hermitian conjugate matrix
$(V^\dag)_{ij}=v^*_{ji}$. Then Eqs.~(\ref{fphi_a}-\ref{fphi_b}) read
\begin{equation} 
 F=V^*V^\tran=F^\dag,\quad \Phi=UV^\tran=\Phi^\tran, 
\label{fphi1} 
\end{equation} 
and Eqs.~(\ref{uv_a}-\ref{uv_b}) can be written as 
\begin{equation} 
\begin{pmatrix} 
U  & V\\ 
V^* & U^* 
\end{pmatrix} 
\begin{pmatrix} 
U^\dag & -V^\tran\\ 
-V^\dag & U^\tran 
\end{pmatrix} 
=\begin{pmatrix} 
\openone & 0\\ 
0& \openone 
\end{pmatrix}, 
\label{uv1} 
\end{equation} 
where $\openone$ denotes the identity matrix. Let us introduce the
composed matrices
\begin{equation} 
\sigma_3=\begin{pmatrix} 
\openone  & 0\\ 
0 & -\openone 
\end{pmatrix},\ 
\sigma_+=\begin{pmatrix} 
\openone  & 0\\ 
0 & 0 
\end{pmatrix}, 
\label{defsig3} 
\end{equation} 
and rewrite Eq.~(\ref{uv1}) in the form 
\begin{equation} 
X\sigma_3X^\dag\sigma_3=\openone, 
\label{uv2} 
\end{equation} 
where $\openone$ stands now for the composed identity matrix, i.e.\ the
r.h.s.\ of Eq.~(\ref{uv1}).  The matrix representation (\ref{uv2}) is
very convenient. For example, from this equation we have
$X^{-1}=\sigma_3 X^\dag\sigma_3$, and
\[ 
\begin{pmatrix} 
\hat{\alpha}\\ 
\hat{\alpha}^\dag 
\end{pmatrix} 
=\sigma_3X^\dag\sigma_3 
\begin{pmatrix} 
\hat{a}\\ 
\hat{a}^\dag 
\end{pmatrix} 
=\begin{pmatrix} 
U^\dag  & -V^\tran\\ 
-V^\dag & U^\tran 
\end{pmatrix} 
\begin{pmatrix} 
\hat{a}\\ 
\hat{a}^\dag 
\end{pmatrix}, 
\] 
which reads in usual notations 
\begin{eqnarray} 
\hat{\alpha}_f&=&\sumpr_\nu\,(u^*_{\nu f}\hat{a}_\nu-v_{\nu f}\hat{a}^\dag_\nu), \nonumber \\ 
\hat{\alpha}^\dag_f&=&\sumpr_\nu\,(u_{\nu f}\hat{a}^\dag_\nu-v^*_{\nu f}\hat{a}_\nu). \nonumber 
\end{eqnarray} 
This equation together with the commutation relations leads to 
\begin{eqnarray} 
\sumpr_\nu\,(u^*_{\nu f}u_{\nu f'}- v_{\nu f}v^*_{\nu f'})&=&\delta(f-f'), \nonumber \\ 
\sumpr_\nu\,(v_{\nu f}u^*_{\nu f'}- u^*_{\nu f}v_{\nu f'})&=&0, \nonumber 
\end{eqnarray} 
which is nothing else but the matrix equation
$X^\dag\sigma_3X\sigma_3=\openone$, resulting from Eq.~(\ref{uv2}).
 
Employing the idea of Ref.~\cite{bogufn}, in which the
Hartree-Fock-Bogoliubov method for Fermi systems was developed, we
define the matrix $K$ with the help of the notations (\ref{fphi1}) and
(\ref{defsig3})
\[ 
K=X^\dag\sigma_3\sigma_+X\sigma_3 
=\begin{pmatrix} 
\openone+F^*  & -\Phi\\ 
\Phi^* & -F 
\end{pmatrix}. 
\] 
Due to Eq.~(\ref{uv2}) and the relation $(\sigma_+)^2=\sigma_+$ we have $K^2=K$. Rewriting the latter equation in terms 
of the matrix $F$ and $\Phi$, we obtain two independent relations 
\begin{eqnarray} 
\Phi^*\Phi&=&F+F^2,
\label{fpfirel_a} \\ 
 F^*\Phi  &=&\Phi F, 
\label{fpfirel_b} 
\end{eqnarray} 
which read in components 
\arraycolsep 0.00 mm 
\begin{equation} 
\sumpr_f\!\Phi^*(f_1,f)\Phi(f,f_2)=\sumpr_f\! F(f_1,f)F(f,f_2)\!+\!F(f_1,f_2),
\label{fpfirel1_a}
\end{equation}
\begin{equation}\sumpr_f\! F(f,f_1)\Phi(f,f_2)=\sumpr_f\!\Phi(f_1,f)F(f,f_2). 
\label{fpfirel1_b} 
\end{equation} 
By using these equations, Eqs.~(\ref{compl_a}-\ref{compl_b}), and the definition $\hat{\vartheta}(x) =\sumpr_\nu\, \hat{a}_\nu
\phi_\nu(x)$,  one can rewrite Eqs.~(\ref{fpfirel1_a}-\ref{fpfirel1_b}) in the coordinate representation
\begin{eqnarray} 
&&\int\d x\, \langle\hat{\vartheta}^\dag(x_1)\hat{\vartheta}^\dag(x)\rangle\langle\hat{\vartheta}(x)
\hat{\vartheta}(x_2)\rangle =\langle\hat{\vartheta}^\dag(x_1)\hat{\vartheta}(x_2)\rangle 
\nonumber \\ 
&&\phantom{\int\d x\, \langle\hat{\vartheta}^\dag} 
+\int\d x\, \langle\hat{\vartheta}^\dag(x_1)\hat{\vartheta}(x)\rangle 
\langle\hat{\vartheta}^\dag(x)\hat{\vartheta}(x_2)\rangle, 
\label{fphicord1}
\end{eqnarray}
\begin{eqnarray} 
&&\int \d x\, \langle\hat{\vartheta}^\dag(x)\hat{\vartheta}(x_1)\rangle\langle\hat{\vartheta}(x)\hat{\vartheta}(x_2)\rangle 
=\int\d x\, \langle\hat{\vartheta}(x_1)\hat{\vartheta}(x)\rangle 
\nonumber\\ 
&&\phantom{\int \d x\, \langle\hat{\vartheta}^\dag(x_1)\hat{\vartheta}(x)\rangle\langle\hat{\vartheta}(x)} 
\times\langle\hat{\vartheta}^\dag(x)\hat{\vartheta}(x_2)\rangle. 
\label{fphicord} 
\end{eqnarray} 
If the condensate depletion is small, one can neglect the second term
in the r.h.s of Eq.~(\ref{fphicord1}), which is of the next
order. Thus, we obtain the expression
\begin{equation}\label{thetaapp} 
\langle\hat{\vartheta}^\dag(x_1)\hat{\vartheta}(x_2)\rangle 
\simeq\int\d x\, \langle\hat{\vartheta}^\dag(x_1)\hat{\vartheta}^\dag(x)\rangle 
\langle\hat{\vartheta}(x)\hat{\vartheta}(x_2)\rangle. 
\end{equation} 
Note that Eq.~(\ref{fphicord}) turns into identity in the
approximation (\ref{thetaapp}), and the same is valid for
Eqs.~(\ref{fpfirel_b}) and (\ref{fpfirel1_b}).



\begin{thebibliography}{99}
\bibitem{classif} We call the potential weak-coupling if it is small and integrable; for more discussions, see
Ref.~\cite{cherny1}.
\bibitem{note1b} In this paper we use the
definition $\xi=\hslash/\sqrt{\mu m}$ for the coherence or healing
length as in Refs.~\cite{shlyap2D,shlyap1D} whereas \cite{dalfovo} uses
$\xi=\hslash/\sqrt{2 \mu m}$.
\bibitem{note2} Due to criterion~(\ref{restr1d}) the expression for
the chemical potential~(\ref{mu1d1}) is not valid formally at very low
densities. However, we arrive at the same result if we calculate first the
interaction energy directly from Eqs.~(\ref{eint1}), (\ref{phiapp1}),
and (\ref{cpar1}), and then integrate over $\lambda$ using the relation
(\ref{varth}).
\bibitem{note3b} The noncondensate fraction was
referred to as ``supracondensate'' in Ref.~\cite{cherny,cherny1,cherny2}.
\end{thebibliography}
\end{document}